\documentclass[12pt]{article}
\usepackage{amsmath}
\usepackage{longtable}
\usepackage{graphicx,psfrag,epsf}
\usepackage{enumerate}
\usepackage{natbib}
\usepackage{chronosys}
\usepackage{setspace}
\usepackage{authblk}
\usepackage{pgfplots}
\pgfplotsset{compat=1.16}
\usepackage{url} 
\newcommand{\blind}{0}
\usepackage{amssymb}
\newcommand{\airbnb}{Airbnb }

\makeatletter
\newcommand*{\addFileDependency}[1]{
  \typeout{(#1)}
  \@addtofilelist{#1}
  \IfFileExists{#1}{}{\typeout{No file #1.}}
}
\makeatother

\catcode`\@=11
\def\chron@selectmonth#1{\ifcase#1\or January\or February\or
March\or April\or May\or June\or July\or August\or
September\or October\or November\or December\fi}

\newcommand{\YoY}{{\text{YoY}}}

\newcommand{\YoYeightteen}{{\text{YoY2018}}}

\title{Lead Times in Flux: Analyzing Airbnb Dynamics During Global Upheavals (2018-2022)}

\setcitestyle{authoryear,open={(},close={)}}
\date{} 
\begin{document}

\def\spacingset#1{\renewcommand{\baselinestretch}%
{#1}\small\normalsize} \spacingset{1}


\if0\blind
{
  \title{\bf LEAD TIMES IN FLUX: ANALYZING AIRBNB BOOKING DYNAMICS DURING GLOBAL UPHEAVALS (2018-2022)}
  \author[1]{Harrison Katz\thanks{\textbf{Corresponding author:} harrison.katz@airbnb.com \\
  \textbf{Phone:} +1 (661) 755-7737 \\
  \textbf{Address:}1641 Veteran Ave, Los Angeles 90024, USA \\
  \textbf{Bio:} Harrison Katz, PhD in Statistics from UCLA, is a Tech Lead Data Scientist on Airbnb’s forecasting team. Previously with the Federal Reserve Board of Governors in risk analysis, his research focuses on financial markets, Bayesian forecasting, and compositional time series.}}
  
  \author[2]{Erica Savage\thanks{\textbf{Email:} erica.savage@airbnb.com \\
  \textbf{Bio:} Erica Savage is an Associate Principal focused on forecasting and
  financial modeling at Airbnb.}}
  
  \author[3]{Peter Coles\thanks{\textbf{Email:} peter.coles@airbnb.com \\
  \textbf{Bio:} Peter Coles is Airbnb’s Head Economist and Director of Data Science. Formerly Head Economist at eBay and a Harvard professor, his research covers marketplace design, friction reduction, data strategy, and public policy. He holds degrees from Princeton and Stanford.}}

  \affil[1]{Data Science, Forecasting, \airbnb}
  \affil[2]{Finance, Forecasting, \airbnb}
  \affil[3]{Data Science, \airbnb}

  \maketitle
} \fi

\if1\blind
{
  \bigskip
  \bigskip
  \bigskip
  \begin{center}
    {\LARGE\bf LEAD TIMES IN FLUX: ANALYZING AIRBNB BOOKING DYNAMICS DURING GLOBAL UPHEAVALS (2018-2022)}
\end{center}
  \medskip
} \fi

\begin{abstract}
\noindent
Short-term changes in booking behaviors can significantly undermine naive forecasting methods in the travel and hospitality industry, especially during periods of global upheaval. Traditional metrics like average or median lead times capture only broad trends, often missing subtle yet impactful distributional shifts. In this study, we introduce a normalized L1 (Manhattan) distance to measure the full distributional divergence in Airbnb booking lead times from 2018 to 2022, with particular emphasis on the COVID-19 pandemic. Using data from four major U.S.\ cities, we find a two-phase pattern of disruption: a sharp initial change at the pandemic’s onset, followed by partial recovery but persistent divergences from pre-2018 norms. Our approach reveals shifts in travelers' planning horizons that remain undetected by conventional summary statistics. These findings highlight the importance of examining the \emph{entire} lead-time distribution when forecasting demand and setting pricing strategies. By capturing nuanced changes in booking behaviors, the normalized L1 metric enhances both demand forecasting and the broader strategic toolkit for tourism stakeholders, from revenue management and marketing to operational planning, amid continued market volatility.
\end{abstract}

\bigskip

\noindent Keywords:  forecasting, lead time distribution, normalized L1 distance, COVID-19 pandemic, Airbnb, tourism analytics, consumer behavior, hospitality, revenue management. 
{\renewcommand{\thefootnote}{}
\footnotetext{ 

}}

\newpage

\doublespacing

\newpage

\section{Introduction}
\label{sec:intro}
Booking lead time—the interval between the booking date and the actual stay date—is a key metric in the travel and hospitality industry. It influences operational and strategic decisions, including pricing strategies, inventory management, marketing campaigns, and customer relationship management. A thorough understanding of booking lead times enables businesses to optimize resource allocation, anticipate demand fluctuations, and enhance competitiveness in a dynamic market environment.

Prior research has underscored the importance of lead times in tourism management. \citet{lim2002time} showed how lead times affect revenue management strategies in the hospitality sector. \citet{tussyadiah2016} examined consumer decision-making processes in tourism, finding that booking lead times are influenced by a complex interplay of factors, including perceived risk, travel purpose, and technological advancements. Lead times are frequently incorporated into forecasts of aggregate demand \citep{fiori2019reservation,pereira2016introduction,zaki2022implementing,fiori2020prediction,webb2022hotel}, while \citet{KATZ20241556} and \citet{de2021lead} presented models for forecasting lead times directly. Moreover, \citet{Song2008} and \citet{Huang2023} have reviewed the broader context of tourism demand forecasting, emphasizing the growing importance of incorporating booking windows into predictive frameworks.

Global events, particularly crises such as pandemics, have profound effects on travel behaviors and lead times. The COVID-19 pandemic, in particular, has led to unprecedented disruptions in the tourism industry \citep{sainaghi2022effects,sharma2021hotels,qiu2020social,mueller2024tourism,okafor2022covid}. According to \citet{gossling2021pandemics}, the pandemic resulted in massive declines in international and domestic travel, altering consumer risk perceptions and booking behaviors. \citet{sigala2020tourism} emphasized that COVID-19 not only impacted demand but also accelerated digital transformation and the adoption of flexible booking policies in the industry. Research suggests that travelers' perceived risk and uncertainty lead to significant changes in booking patterns. For example, \citet{lo2011hong} found that during health crises, travelers tend to delay bookings or opt for destinations perceived as safer. \citet{fuchs2011exploratory} showed that risk reduction strategies, such as flexible cancellation policies and enhanced safety measures, become more critical in influencing booking lead times. 

Additionally, recent research in the airline industry underscores similar shifts in booking lead-time patterns under COVID-19. \citet{zhang2021impact} investigated the Chinese aviation market and revealed how pandemic-related uncertainties drove travelers to shorten planning horizons and switch travel modes. Their moment-based analysis demonstrates that variance and higher-order distributional features of lead times can provide actionable insights into passenger behavior, a perspective that aligns well with the comprehensive approach advocated in this paper.

Despite recognition of these shifts, much of the existing literature relies on traditional statistical measures—such as mean and median lead times—to analyze booking behaviors. While these measures are useful, they may not fully capture the distributional changes and increased variability introduced by external shocks like pandemics. As \citet{ASSAF201943,ASSAF2019582,assaf2019quantitative} argued, there is a need for more advanced analytical approaches in tourism research to better understand complex consumer behaviors. Likewise, \citet{Assaf2019c} underscored the importance of distribution-aware metrics in examining tourism demand, especially during times of heightened uncertainty.

This study addresses this gap by introducing the normalized Manhattan (L1) distance metric as a novel approach to analyzing variations in booking lead time distributions. Unlike traditional metrics, the normalized L1 distance captures changes across the entire distribution, providing a more sensitive and comprehensive measure of divergence in booking behaviors over time. This method allows for the detection of shifts that might be overlooked when relying solely on averages or standard deviations. 

The primary objective of this research is to assess the effectiveness of the normalized L1 distance metric in detecting and quantifying changes in Airbnb booking lead times during periods of global upheaval, specifically the COVID-19 pandemic. By analyzing data from Airbnb properties across four major U.S. urban centers—Austin, Boston, Miami, and San Francisco—from 2018 to 2022, we aim to provide a comprehensive perspective on the evolving dynamics of urban tourism. Our work also resonates with the emerging body of research on digital platforms’ responses to crises; as noted by \citet{tussyadiah2016} and \citet{Gyodi2021}, peer-to-peer accommodation providers (like Airbnb) can exhibit different booking behaviors compared to traditional hotels, a gap that demands further empirical and methodological attention.

Furthermore, this study contributes to the literature by offering empirical evidence on how external shocks impact booking behaviors in the sharing economy context. By situating our findings within the broader theoretical frameworks of risk perception and consumer behavior during crises, we enhance the understanding of how travelers adjust their planning in response to heightened uncertainty. Our focus on Airbnb properties answers recent calls \citep[e.g.,][]{Sigala2020} to examine how platform-specific features—such as flexible listing types, decentralized inventory, and digital-first user experiences—shape travel demand under extreme market disruptions.

The normalized L1 distance is used to quantify the divergence in lead time distributions across different years, offering a non-parametric measure that adjusts for scale differences by normalizing the traditional L1 (Manhattan) distance. This approach gives a global summary of lead time distributional changes, allowing a more nuanced understanding of how booking patterns have shifted in response to significant events. This framework provides practical insights for industry stakeholders and also advances methodological approaches in tourism research, as it addresses the need for enhanced analytics that capture full distributional variations rather than focusing on singular statistical descriptors alone.

\section{Methodology}
\label{sec:methodology}

This section describes our approach to studying changes in Airbnb booking lead-time distributions from 2018 to 2022. In particular, we define the key variable of interest (\S\ref{subsec:data}), introduce the normalized L1 (Manhattan) distance for quantifying distributional divergences (\S\ref{subsec:l1_metric}), and outline our time-series decomposition and interpretative strategy (\S\ref{subsec:stl_decomp}).

\subsection{Data and Variable of Interest}
\label{subsec:data}

We analyze Airbnb booking data for four major U.S.\ cities (Austin, Boston, Miami, and San Francisco) from January~2018 through December~2022. For each city, we consider both \emph{destination} bookings (nights stayed \emph{in} the city) and \emph{origin} bookings (nights booked \emph{from} the city), distinguishing between domestic and international travel corridors. In total, the dataset comprises monthly aggregates of booking records, where each record indicates how many nights were booked in a given city (or from a given city) for specific calendar months of stay.

A key variable in our study is the \emph{lead time}, defined as the number of days between the booking date and the check-in date. We confine our analysis to lead times from 0 to 365 days before check-in. Although travelers \emph{can} book more than one year in advance, these cases represent under .5\% of total bookings in our dataset and are thus negligible. By restricting the lead-time domain to 0--365 days, we ensure that more than 99\% of all nights are captured, while also simplifying the analysis (e.g., a uniform bin structure for the lead-time distribution).

\paragraph{Monthly Lead-Time Distributions.}
Let \(B_t\) be the set of all bookings whose check-in date (i.e., first night) falls in month~\(t\). 
For each booking \(i \in B_t\),
\begin{itemize}
  \item \(\Delta_i \in \{0,1,\ldots,365\}\) denotes the lead time (in days) between its booking date and its check-in date,
  \item \(\ell_i\) denotes the number of nights in booking~\(i\).
\end{itemize}
For each lead time \(\Delta\in\{0,\dots,365\}\), define
\[
  w_t(\Delta) 
  \;=\; 
  \sum_{\substack{i \in B_t \\ \Delta_i = \Delta}} 
  \ell_i,
\]
that is, the total number of \emph{nights} contributed by all bookings in \(B_t\) whose lead time is exactly \(\Delta\).\footnote{%
  For example, if two guests each book 5-night stays with \(\Delta_i=3\), then they add \(5+5=10\) to \(w_t(3)\).}

Since \(\Delta\) ranges from 0 to 365, we have \(366\) possible lead-time categories. To convert these totals into a \emph{probability distribution}, we normalize:
\begin{equation}
\label{eq:L_tDelta}
  L_{t}(\Delta) 
  \;=\; 
  \frac{w_t(\Delta)}{\sum_{\delta=0}^{365} w_t(\delta)},
  \quad 
  \Delta = 0,1,\ldots,365.
\end{equation}
Hence, \(L_{t}(\Delta)\) represents the \emph{proportion} of all nights (among bookings that start in month~\(t\)) that were booked \(\Delta\) days in advance. By construction, 
\[
  \sum_{\Delta=0}^{365} L_{t}(\Delta) 
  \;=\; 
  1.
\]

\subsection{Descriptive Analysis}
A descriptive analysis is performed for each city, examining each year separately. We look at:
\begin{itemize}
    \item \textbf{Mean trip lead times}: The mean trip lead time provides the average number of days between the booking date and the trip date, weighted by the trip length. It gives an overall sense of how far in advance travelers are planning their trips. The mean is calculated using the formula:

    \[
    \text{Mean} = \frac{1}{n} \sum_{i=1}^{n} x_i
    \]

    where \( n \) is the total number of trips and \( x_i \) is the lead time for each trip.

    \item \textbf{Median trip lead times}: The median trip lead time indicates the central tendency of trip lead times, offering a measure less sensitive to extreme outliers compared to the mean. This is particularly useful for understanding typical trip planning behavior in the presence of significant variability. It is calculated as follows:

    \begin{itemize}
        \item If \( n \) (the number of observations) is odd, the median is the middle value:

        \[
        \text{Median} = x_{\left(\frac{n+1}{2}\right)}
        \]

        \item If \( n \) is even, the median is the average of the two middle values:

        \[
        \text{Median} = \frac{x_{\left(\frac{n}{2}\right)} + x_{\left(\frac{n}{2} + 1\right)}}{2}
        \]

        where \( x_{(i)} \) represents the \( i \)-th ordered lead time.
    \end{itemize}

    \item \textbf{Standard deviation of trip lead times}: The standard deviation measures the variability in trip lead times, indicating how spread out the planning patterns are around the mean. This statistic helps to assess the consistency of trip planning behaviors over time. The standard deviation is computed as:

    \[
    \text{Standard Deviation} = \sqrt{\frac{1}{n-1} \sum_{i=1}^{n} (x_i - \text{Mean})^2}
    \]

    where \( x_i \) is the lead time for each trip, and the mean is as defined above.

\end{itemize}

\subsection{Divergence Analysis Using Normalized L1 Distance}
\label{subsec:l1_metric}

While basic summary statistics (mean, median, and standard deviation) can highlight overall trends in lead times, they may overlook \emph{distributional shifts} that arise during periods of upheaval. To address this, we employ the \emph{normalized L1 distance} (a.k.a.\ the Manhattan distance).

\paragraph{Intuition:} 
The L1 distance between two probability distributions \(p\) and \(q\) on the same domain (in our case, \(\Delta = 0,1,\dots,365\)) essentially sums the absolute differences of their probabilities at each point:
\[
  D_{\text{L1}}(p,q) \;=\; \sum_{\Delta=0}^{365}|p(\Delta) - q(\Delta)|.
\]
If the distributions do not overlap at all (one is \(1\) in places where the other is \(0\), and vice versa), the L1 distance is maximal. Because we want a measure in \([0,1]\), we divide by 2:
\[
 D(p,q) \;=\; \frac{1}{2} \sum_{\Delta=0}^{365}|p(\Delta) - q(\Delta)|.
\]
Interpretation: \(D(p,q)\) is the fraction of the total probability mass that has “shifted.” A value of \(0\) means no difference (identical distributions), and \(1\) implies total divergence.

\begin{figure}[ht]
\centering
\includegraphics[width=1.15\textwidth]{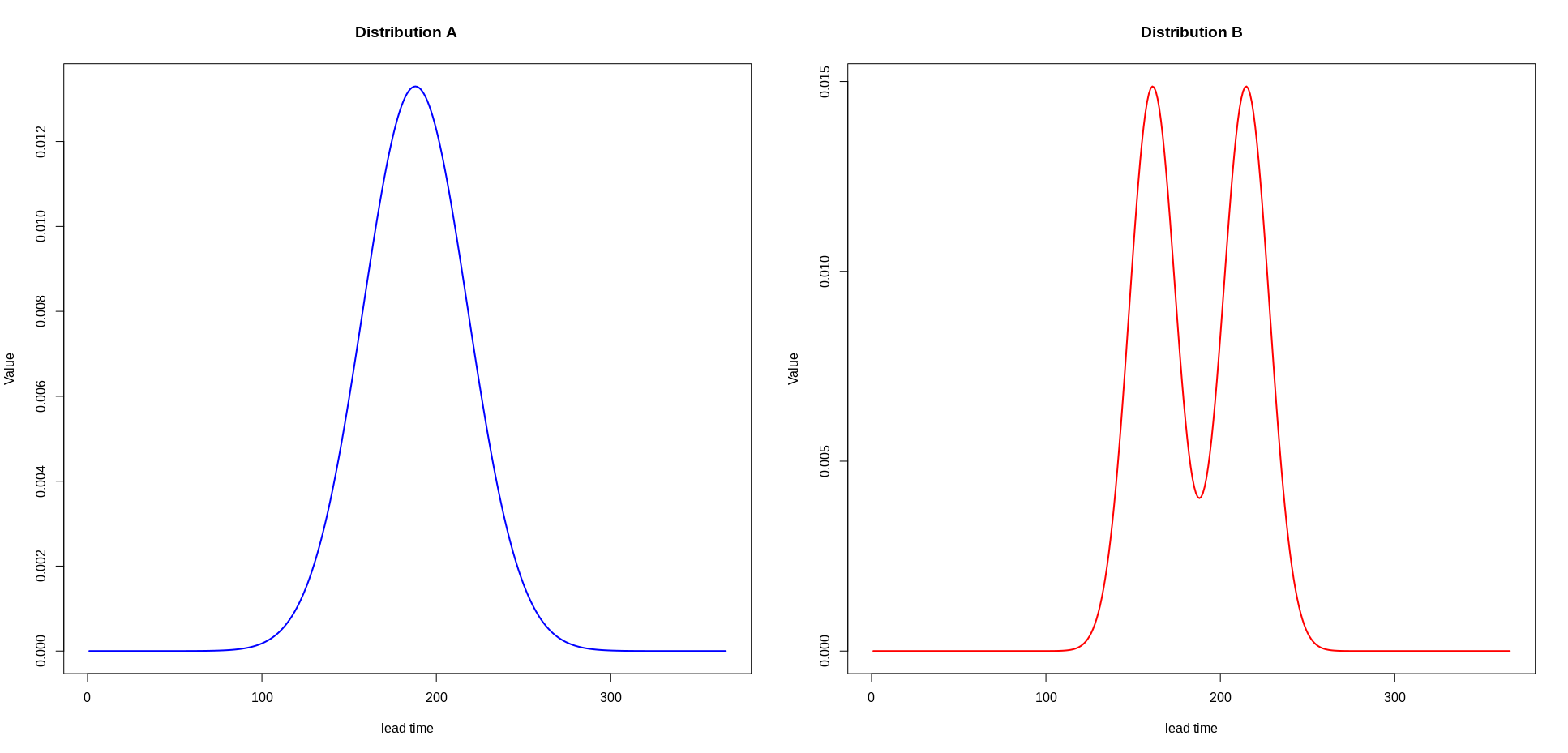}\
\caption{\emph{Example of two lead-time distributions, A (left) and B (right), that share the same mean ($\mu=188$) and standard deviation ($\sigma=30$), but exhibit notably different shapes. Distribution~A is roughly unimodal, while Distribution~B is bimodal. Despite having identical mean and SD, the two distributions have an L1 distance of $0.25$, highlighting how summary statistics alone may fail to capture important shifts.}}
\label{fig:distA}
\end{figure}

\noindent
Figure~\ref{fig:distA} illustrates why L1 distance can be more informative. Distributions~A and~B both have mean~\(188\) and standard deviation~\(30\), yet differ substantially in shape (unimodal vs.\ bimodal). Commonly used summary statistics do not reveal these differences, whereas the normalized L1 distance reveals a notable shift: here, \(D(A,B) = 0.25\), indicating that 25\% of the probability mass has effectively “moved.”

\subsubsection{Year-over-Year vs.\ Baseline Comparisons}
\label{subsec:comparisons}
We calculate the normalized L1 distance between lead-time distributions in two ways:

\begin{enumerate}
    \item \textbf{Year-over-Year (YoY) Comparison:} 
    \[
      l^1_{\YoY,t} \;=\; \frac{1}{2} \sum_{\Delta=0}^{\Delta_{\max}} 
            |L_{t}(\Delta) - L_{t-12}(\Delta)|.
    \]
    This measures how much the distribution in month \(t\) diverges from the same calendar month in the previous year.
    
    \item \textbf{Baseline-2018 Comparison:} 
    \[
      l^1_{\YoYeightteen,t} \;=\; \frac{1}{2} \sum_{\Delta=0}^{\Delta_{\max}} 
            |L_{t}(\Delta) - L_{2018,t}(\Delta)|,
    \]
    where \(L_{2018,t}(\Delta)\) denotes the distribution for the corresponding month of 2018. This approach captures how far month \(t\) is from a pre-pandemic benchmark.
\end{enumerate}

These two perspectives reveal distinct patterns. The YoY metric highlights sequential shifts (e.g., from 2020 to 2021), while the Baseline-2018 metric clarifies whether distributions are returning to pre-crisis norms or have settled into new equilibria.

\subsection{Seasonal-Trend Decomposition (STL)}
\label{subsec:stl_decomp}

After computing the L1 distances over time, we apply a Seasonal-Trend decomposition using Loess (STL) \citep{cleveland1990stl} to each L1 time series. The STL method splits each series \( y_t \) into three components:
\[
   y_t = T_t + S_t + R_t,
\]
where \(T_t\) is a slowly varying trend, \(S_t\) is a seasonal component capturing recurring monthly or weekly patterns, and \(R_t\) is a remainder (irregular fluctuations). We prefer STL for its:

\begin{itemize}
    \item \textbf{Robustness to Outliers:} It uses iterative Loess smoothing to handle sudden spikes or dips.  
    \item \textbf{Flexibility in Seasonal Patterns:} Each city may show distinct seasonalities (e.g., academic schedules, major events).  
    \item \textbf{Ease of Interpretation:} We can pinpoint the long-term shift in lead-time divergences by examining \(T_t\).
\end{itemize}

Using STL on \(l^1_{\YoY,t}\) and \(l^1_{\YoYeightteen,t}\) helps identify whether divergence spikes are driven mainly by structural (long-term) changes or by seasonal and event-driven fluctuations.

\section{Toy Example: Impact of Lead Time Changes on Naive Forecast Methods}

\subsection{Setup}
To illustrate the sensitivity of the normalized L1 metric to shifts in lead time distributions and to demonstrate how changes in lead times can affect commonly used forecast methods, we present a simulated example using a fictional town called B-Ville. In this example, we assume that in the year 2019, B-Ville had 1,000 total bookings for a specific trip date, and in 2020, it had 1,200 total bookings for the same trip date.

\begin{figure}
    \includegraphics[scale=.6]{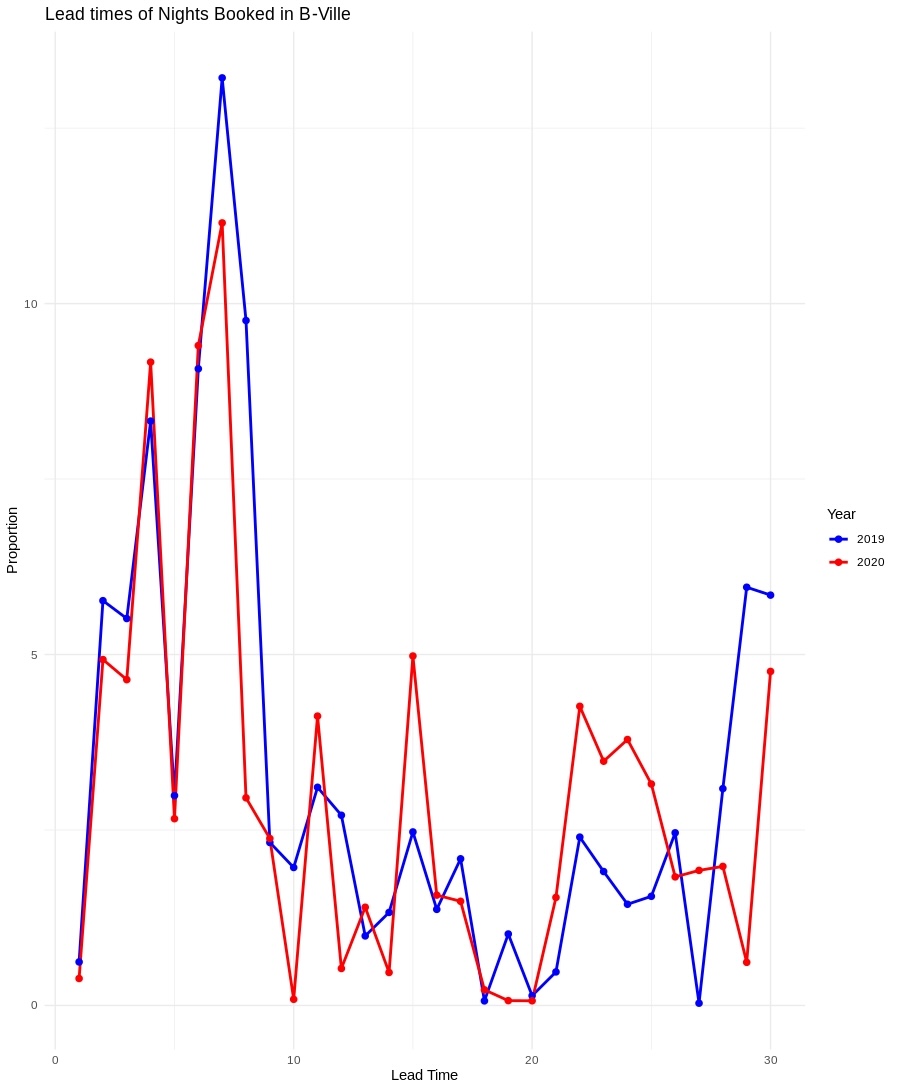}
    \caption{The distribution of lead times for the fictional town B-Ville, illustrating the changes in booking behaviors across two consecutive years. This figure shows how the normalized L1 distance metric captures small perturbations in lead times better than traditional summary statistics like mean, median, or standard deviation, thereby emphasizing its sensitivity to distributional changes in booking behaviors.}
        \label{toy_example}
\end{figure}

We created a lead time distribution—a compositional vector of size 30—representing the proportion of bookings made at each lead time from 1 to 30 days before the trip date. To simulate changes in booking behaviors from 2019 to 2020, we introduced a small perturbation to the 2019 lead time distribution by adding random noise drawn from a half-normal distribution with scale parameter $\sigma=0.05$ to each component and then re-normalizing the vector to ensure it sums to one. This resulted in the 2020 lead time distribution, reflecting a shift in booking patterns. The lead time distributions for both years are shown in Figure~\ref{toy_example}.

To estimate the pickup curves, we calculated the cumulative sum of the lead time distributions for each year. The pickup curves represent the accumulation of bookings as the trip date approaches, showing the cumulative percentage of total bookings made up to each lead time. 

\subsection{Results and Effects on the Pickup Method}

\begin{figure}[ht]
    \centering
    \includegraphics[scale=0.65]{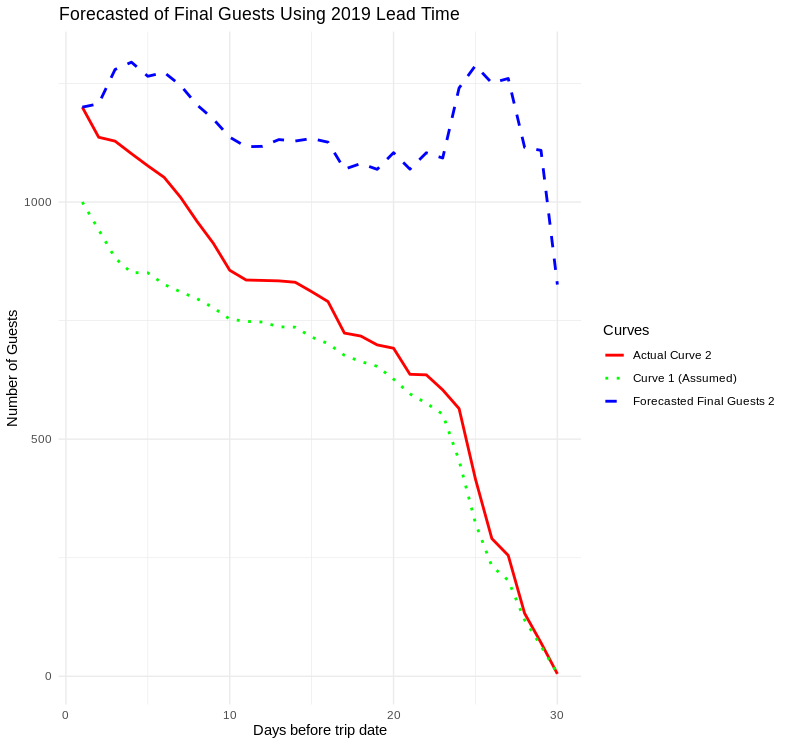}
    \caption{Pickup curves (cumulative sum of lead time distributions) for B-Ville in 2019 and 2020. The 2020 pickup curve indicates a slower accumulation of bookings compared to 2019, leading to forecasting errors when using the pickup method.}
    \label{toy_example_pickup}
\end{figure}

The pickup method\footnote{In this example, we apply the pickup method under the assumption of unconstrained demand. This means we consider that supply limitations do not affect the total number of bookings, allowing demand to be fully realized.} is a common forecasting technique \citep{pereira2016introduction} in the hospitality industry, where future bookings are predicted based on historical booking patterns and current observed bookings. For instance, suppose we are 17 days before the trip date in 2020 and have recorded 723 bookings so far. In 2019, the pickup curve indicates that by 17 days before the trip date, 69\% of bookings had been made. Applying the multiplicative pickup method, we would predict the total number of bookings for 2020 to be

\begin{equation}
\text{Forecasted Bookings} = \frac{\text{Current Bookings}}{\text{Cumulative \% at $t=17$ (from 2019 data)}} = \frac{723}{0.69} \approx 1,047.
\label{eq:forecasted_bookings}
\end{equation}
However, the actual total bookings in 2020 are 1,200. This means that the forecasted bookings underestimate the actual bookings by approximately 150 bookings, or 12.5\%. This discrepancy arises because the lead time distribution shifted in 2020, with more bookings occurring closer to the trip date—a change captured by the normalized L1 distance between the 2019 and 2020 lead time distributions, which is 0.2156. This is illustrated in Figure~\ref{toy_example_pickup}, which displays both the historical pickup curves and the forecasted total bookings at each time point.

Traditional summary statistics fail to capture this change effectively. As shown in Table~\ref{tab:toy_example}, the mean lead time increased by only 3.58\%, the median remained unchanged, and the standard deviation decreased by 4.30\%. These minor changes in conventional metrics mask the significant distributional shift affecting the pickup method's forecast.

\begin{table}[ht]
\centering
\begin{tabular}{lccc}
  \hline
  & Year 1 & Year 2 & Change (\%) \\
  \hline
  Mean & 12.38066 & 12.82447 & 3.58 \\
  Median & 8 & 8 & 0 \\
  Standard Deviation & 9.417028 & 9.012016 & -4.30 \\
  L1 Distance Change & \multicolumn{2}{c}{0.2156462} & \\
  \hline
\end{tabular}
\caption{Results from simulated lead times data in B-Ville. The L1 distance change better captures the distributional shifts compared to mean, median, or standard deviation alone.}
\label{tab:toy_example}
\end{table}

This example illustrates how changes in lead time distributions, as quantified by the normalized L1 metric, can impact the accuracy of forecasts made using historical lead times. When lead time distributions shift, relying solely on historical booking patterns may lead to underestimation or overestimation of future bookings.

\subsection{Bounding the Forecast Error of Naive Methods Using the Normalized L1 Distance}

We aim to derive a bound on the relative forecast error \( \epsilon \) of the pickup method that incorporates both the normalized L1 distance \( D \) between the historical and actual lead time distributions and the time remaining until the trip date.

Let:

\begin{itemize}
    \item \( L_{\text{hist}}(k) \) and \( L_{\text{actual}}(k) \) denote the historical and actual lead time distributions, where \( k \) ranges from 1 to \( \Delta_{\text{max}} \) (the maximum lead time considered).
    \item \( C_{\text{hist}}(\Delta) = \sum_{k=1}^{\Delta} L_{\text{hist}}(k) \) and \( C_{\text{actual}}(\Delta) = \sum_{k=1}^{\Delta} L_{\text{actual}}(k) \) are the cumulative distributions up to lead time \( \Delta \).
    \item \( D = \frac{1}{2} \sum_{k=1}^{\Delta_{\text{max}}} |L_{\text{hist}}(k) - L_{\text{actual}}(k)| \) is the normalized L1 distance between the distributions.
    \item \( \Delta_\text{max} \) is the maximum lead time considered (e.g., 365 days).
\end{itemize}

The \textbf{relative forecast error} \( \epsilon \) of the pickup method is defined as:

\begin{equation}
\epsilon = \frac{\hat{B}_{\text{total}} - B_{\text{total}}}{B_{\text{total}}}
\label{eq:relative_forecast_error}
\end{equation}

where:

\begin{itemize}
    \item \( \hat{B}_{\text{total}} \) is the forecasted total number of bookings using the pickup method.
    \item \( B_{\text{total}} \) is the actual total number of bookings.
\end{itemize}

Using the pickup method, the forecasted total bookings are calculated as:

\[
\hat{B}_{\text{total}} = \frac{B_{\text{obs}}}{C_{\text{hist}}(\Delta)}
\]

where:

\begin{itemize}
    \item \( B_{\text{obs}} \) is the number of bookings observed up to lead time \( \Delta \).
    \item \( C_{\text{hist}}(\Delta) \) is the cumulative historical booking proportion up to lead time \( \Delta \).
\end{itemize}

At time \( \Delta \), the actual cumulative booking proportion is:

\[
C_{\text{actual}}(\Delta) = \frac{B_{\text{obs}}}{B_{\text{total}}}
\]

Substituting these expressions into Equation~\ref{eq:relative_forecast_error}, we derive (see Appendix~\ref{appendix:derivation}) the following bound on the relative forecast error \( \epsilon \):

\begin{equation}
|\epsilon| \leq \frac{2D \left( 1 - \frac{\Delta}{\Delta_\text{max}} \right)}{C_{\text{hist}}(\Delta)}
\label{eq:error_bound_improved}
\end{equation}

This bound decreases as \( \Delta \) increases (i.e., as we approach the trip date) and approaches zero when \( \Delta = \Delta_\text{max} \).

\subsubsection{Estimating the Normalized L1 Distance \( D \)}

It is important to note that \( D \) represents the divergence between the historical and actual lead time distributions, which is unknown in real time because the actual distribution \( L_{\text{actual}}(k) \) is not fully observed until after the trip date. However, practitioners can estimate \( D \) by:

\begin{itemize}
    \item \textbf{Historical Analysis}: Examining historical patterns of divergence during similar periods or events to estimate a typical or maximum value of \( D \).
    \item \textbf{Scenario Analysis}: Considering specific values of \( D \) based on past maximum observed divergences to model potential forecast errors under different scenarios.
    \item \textbf{Real-Time Monitoring}: Tracking early indicators of changes in booking behaviors (e.g., sudden shifts in booking pace) to adjust the estimated \( D \).
\end{itemize}

By using these methods, forecasters can approximate \( D \) and apply the bound in Equation~\ref{eq:error_bound_improved} to assess the potential forecast error under various scenarios.

\subsubsection{Example}

Revisiting the toy example of B-Ville, suppose we estimate \( D \) based on historical divergences:

\begin{itemize}
    \item \textbf{Estimated Normalized L1 Distance}: \( D = 0.2156 \) (based on a previously observed maximum divergence).
    \item \textbf{Maximum Lead Time}: \( \Delta_{\text{max}} = 30 \) days.
    \item \textbf{At \( \Delta = 17 \) days} (13 days before the trip date):
        \begin{align*}
        1 - \frac{\Delta}{\Delta_{\text{max}}} &= 1 - \frac{17}{30} = \frac{13}{30} \\
        C_{\text{hist}}(17) &= 0.69 \\
        |\epsilon| &\leq \frac{0.2156 \times 2 \times \frac{13}{30}}{0.69} \approx 0.27
        \end{align*}
    \item \textbf{At \( \Delta = 25 \) days} (5 days before the trip date):
        \begin{align*}
        1 - \frac{\Delta}{\Delta_{\text{max}}} &= 1 - \frac{25}{30} = \frac{5}{30} = \frac{1}{6} \\
        C_{\text{hist}}(25) &= 0.90 \\
        |\epsilon| &\leq \frac{0.215 \times 2 \times \frac{1}{6}}{0.90} \approx 0.08
        \end{align*}
\end{itemize}

This example demonstrates how \( D \) allows practitioners to use the bound to assess potential forecast errors at different lead times.

\paragraph{Note}
The detailed derivation of this bound is provided in Appendix~\ref{appendix:derivation}, where we explain how it arises from the properties of cumulative distributions and the normalized L1 distance.

\section{Airbnb Results}
In this section, we present both conventional summary metrics (mean, median, standard deviation) and the normalized L1 divergence for booking lead times across four U.S.\ cities (Austin, Boston, Miami, and San Francisco) from 2018 to 2022. Our aim is to show how these two approaches \emph{complement} each other in revealing shifts in booking behavior—while mean or median changes capture overall trends in travelers' planning horizons, the L1 metric uncovers subtle distributional movements that may not be reflected in central tendencies.

\subsection{Evolution of Lead Times}

\begin{figure}
    \includegraphics[scale=.75]{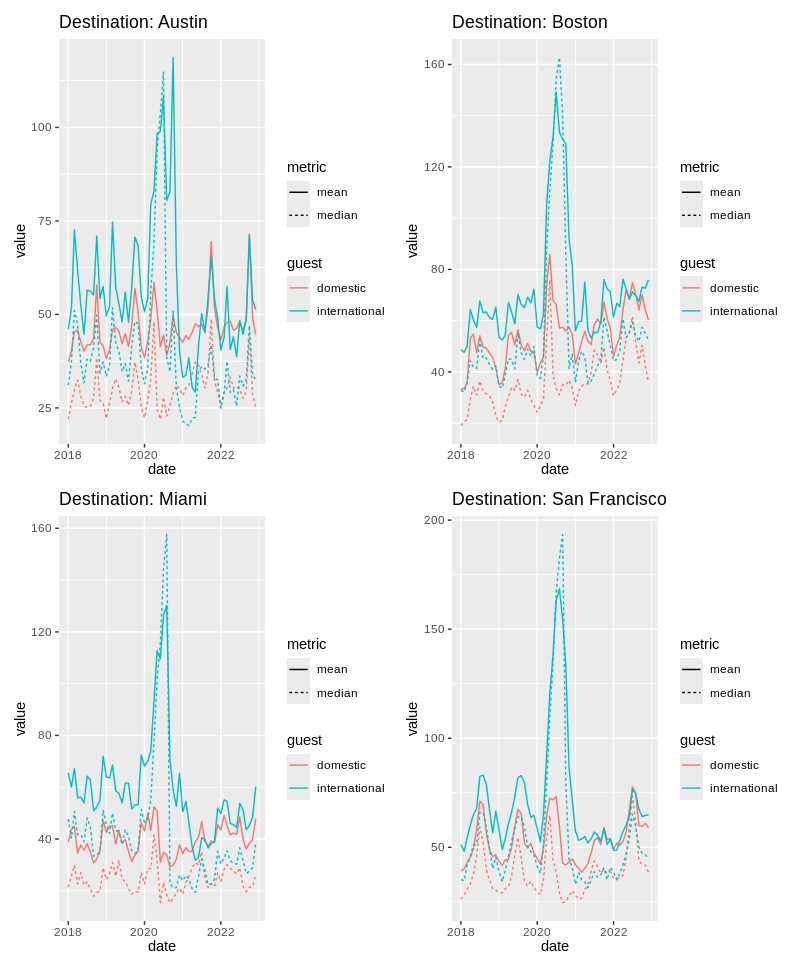}
    \caption{Trends in mean (solid line) and median (dashed line) lead times for Airbnb bookings to Austin, Boston, Miami, and San Francisco from 2018 to 2022. The figure differentiates domestic (red) and international (blue) travel, showing how booking behaviors evolved over time, particularly during significant events such as the COVID-19 pandemic.}
        \label{mean_median_trip_destination}
\end{figure}

\begin{figure}
    \includegraphics[scale=.75]{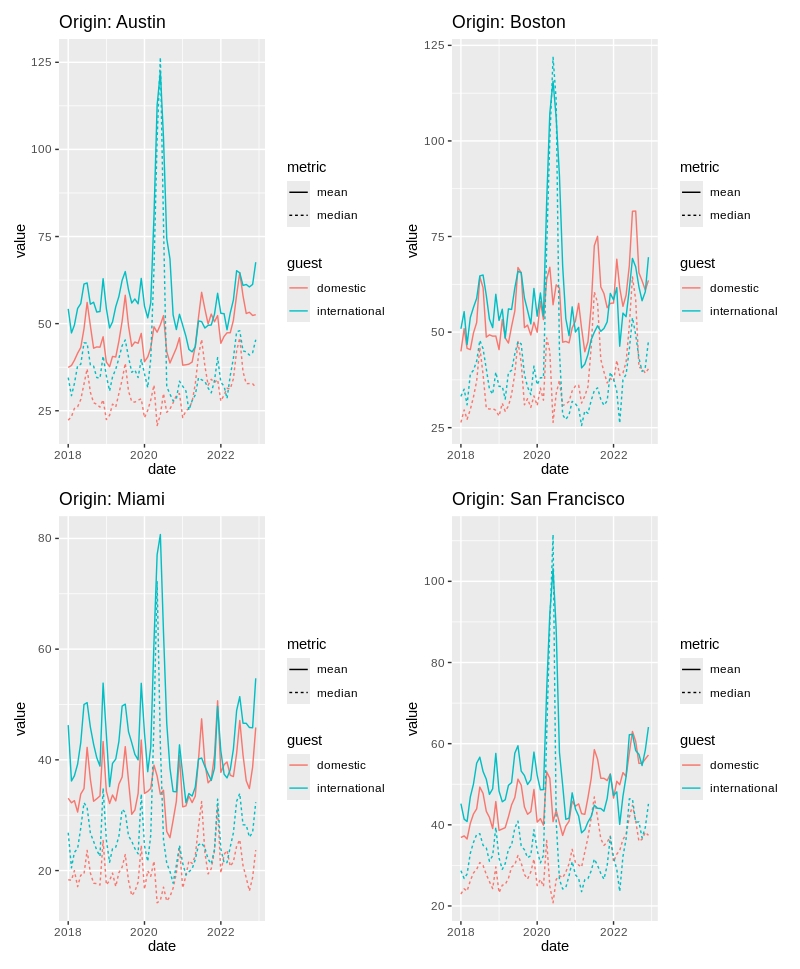}
    \caption{Trends in mean (solid line) and median (dashed line) lead times for Airbnb bookings from Austin, Boston, Miami, and San Francisco from 2018 to 2022. The figure differentiates domestic (red) and international (blue) travel, showing how booking behaviors evolved over time, particularly during significant events such as the COVID-19 pandemic. }
        \label{mean_median_trip_origin}
\end{figure}

The analysis of Airbnb booking data from 2018 to 2022 shows significant variations in lead times across the cities of Austin, Boston, Miami, and San Francisco. We examined the mean, median, and standard deviation (SD) of lead times, segmented by domestic and international travel, and categorized by origin (bookings from the city) and destination (bookings to the city). Figures \ref{mean_median_trip_destination} and \ref{mean_median_trip_origin} show the trends in mean and median lead times, while Supplementary Tables \ref{yearly_summary_domestic} and \ref{yearly_summary_international} provide detailed yearly summaries.

\subsubsection{2018--2019: Pre-Pandemic Period}

During the pre-pandemic period, Miami consistently had the shortest median lead times for both inbound and outbound domestic travel, suggesting a higher propensity for last-minute bookings among travelers to and from Miami. For international travel, bookings originating from Miami and those destined for Austin had the shortest median lead times. In contrast, San Francisco had the longest median lead times for domestic and international destination bookings, indicating that travelers to San Francisco tended to plan their trips further in advance. On an origin basis, Boston had the longest median lead times for both domestic and international bookings, suggesting that residents of Boston planned their trips earlier than those from other cities.

\subsubsection{2020: Impact of the COVID-19 Pandemic}

The onset of the COVID-19 pandemic in 2020 caused significant disruptions in travel patterns, particularly in international travel. International destination median lead times surged by at least 20\% in every city, with Boston and San Francisco experiencing the most substantial increases of 64\% and 42\%, respectively. The SD of lead times also reached its highest levels for international travel in 2020, reflecting heightened variability in booking behaviors.

Domestic travel had mixed responses. While Boston's domestic destination median lead times increased by 17\%, both San Francisco and Miami experienced decreases of 11\% and 8\%, respectively. This suggests that travelers to certain cities reacted differently to the pandemic's onset. The SD for domestic travel lead times increased in several cities, indicating greater variability amid uncertainty.

\subsubsection{2021: Recovery and Shifts in Booking Behavior}

In 2021, as travel began to rebound, domestic lead times increased across all cities. Notably, Miami's origin median lead time surged by 35\%, and Austin's destination median lead time increased by 21\%. This suggests a return to more advanced planning for domestic trips. For international travel, lead times generally decreased in most cities, possibly reflecting pent-up demand and travelers' eagerness to book trips as restrictions eased. San Francisco and Miami saw significant reductions in destination median lead times by 48\% and 52\%, respectively. Boston's origin median lead time for international travel decreased by 29\%.

\subsubsection{2022: Stabilization and Continued Variability}

By 2022, changes in mean and median lead times for domestic travel were generally less pronounced, suggesting a stabilization of booking behaviors. However, San Francisco experienced a notable 28\% increase in destination median lead times, indicating that travelers to San Francisco were planning trips even further in advance than before. The SD reached its highest level for all origin domestic travel in 2022, implying persistent variability.

For international travel, mean and median lead times increased again for all cities, reflecting a cautious return to international travel with longer planning horizons. San Francisco experienced the largest year-over-year increases in origin mean and median lead times at 27\% and 31\%, respectively. Similarly, bookings to Miami saw median lead times increase by 23\% and mean lead times by 22\%.

\clearpage

\subsection{Divergence Analysis Using Normalized L1 Distance}

\begin{figure}
    \includegraphics[scale=.75]{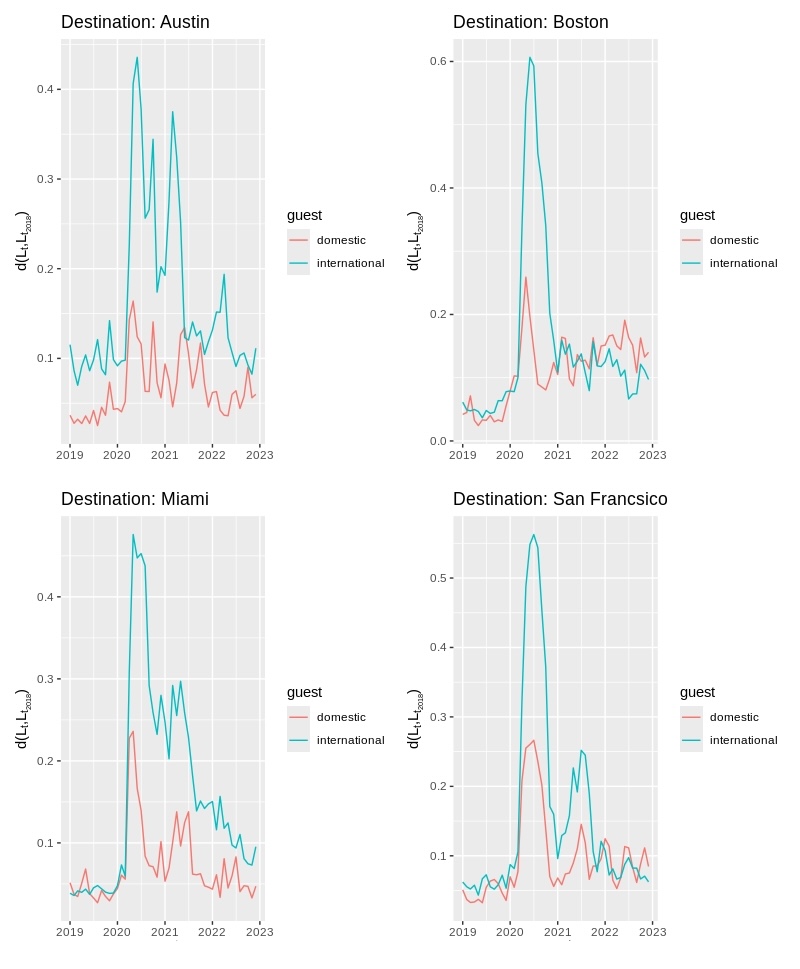}
    \caption{Normalized L1 divergence in \textbf{destination} lead times for Airbnb bookings, taking 2018 as the baseline year. }
        \label{l1_yo2018_destination}
\end{figure}

\begin{figure}
    \includegraphics[scale=.75]{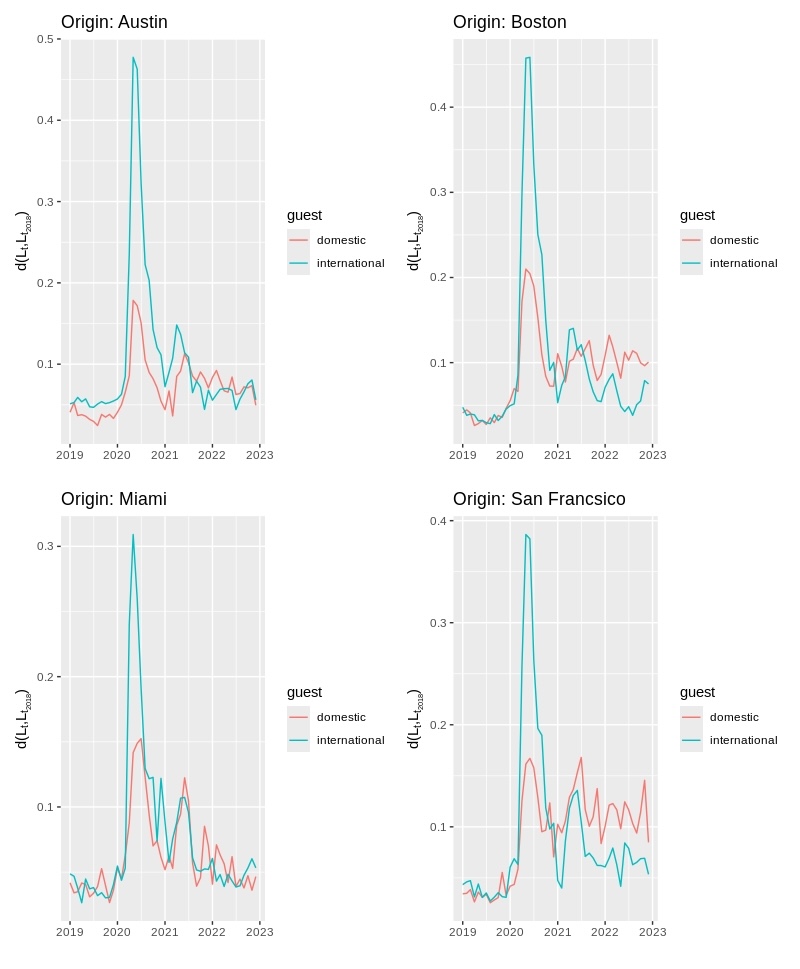}
    \caption{Normalized L1 divergence in \textbf{origin} lead times for Airbnb bookings, taking 2018 as the baseline year.}
        \label{l1_yo2018_origin}
\end{figure}

To capture shifts across the \emph{entire} lead-time distribution, we use the normalized L1 distance, calculated both relative to the baseline year 2018 ($l^1_{\YoYeightteen,t}$) and relative to the same month in the previous year ($l^1_{\YoY,t}$). Figures~\ref{l1_yo2018_destination}--\ref{l1_yo2018_origin} depict the baseline comparisons, while Figures~\ref{l1_yoy_destination}--\ref{l1_yoy_origin} illustrate the year-over-year divergences. The associated numeric summaries appear in Supplementary Tables~\ref{l1_domestic_yoy}--\ref{l1_international_2018}.

\subsubsection{Baseline Comparison Divergence ($l^1_{\YoYeightteen,t}$)}

Comparisons with 2018 underscore how dramatically the pandemic reshaped booking behaviors. For international travel, $l^1_{\YoYeightteen,t}$ peaked in 2020 across all four cities, indicating substantial distributional disruption. In Boston, for instance, the median divergence for international destination bookings rose from 0.05 in 2019 to 0.34 in 2020, while in Austin it climbed from 0.05 to 0.17. Although these divergences eased somewhat in 2021 and 2022, they generally remained elevated compared to pre-pandemic levels, suggesting that international booking patterns never fully reverted to their 2018 norms.

A similar story emerges for domestic bookings: San Francisco’s median $l^1_{\YoYeightteen,t}$ increased from 0.03 in 2019 to 0.11 in 2020 and stabilized around 0.11--0.12 thereafter, while Boston’s divergence rose from 0.03 to 0.10 and ultimately reached 0.15 by 2022. These sustained gaps indicate that, even when the mean or median domestic lead time did not exhibit a massive year-to-year swing, \emph{parts} of the distribution migrated toward notably shorter or longer booking windows. As a result, the overall distribution in 2022 still looked markedly different from its 2018 baseline.

\clearpage

\subsubsection{Year-over-Year Divergence ($l^1_{\YoY,t}$)}

\begin{figure}
    \includegraphics[scale=.85]{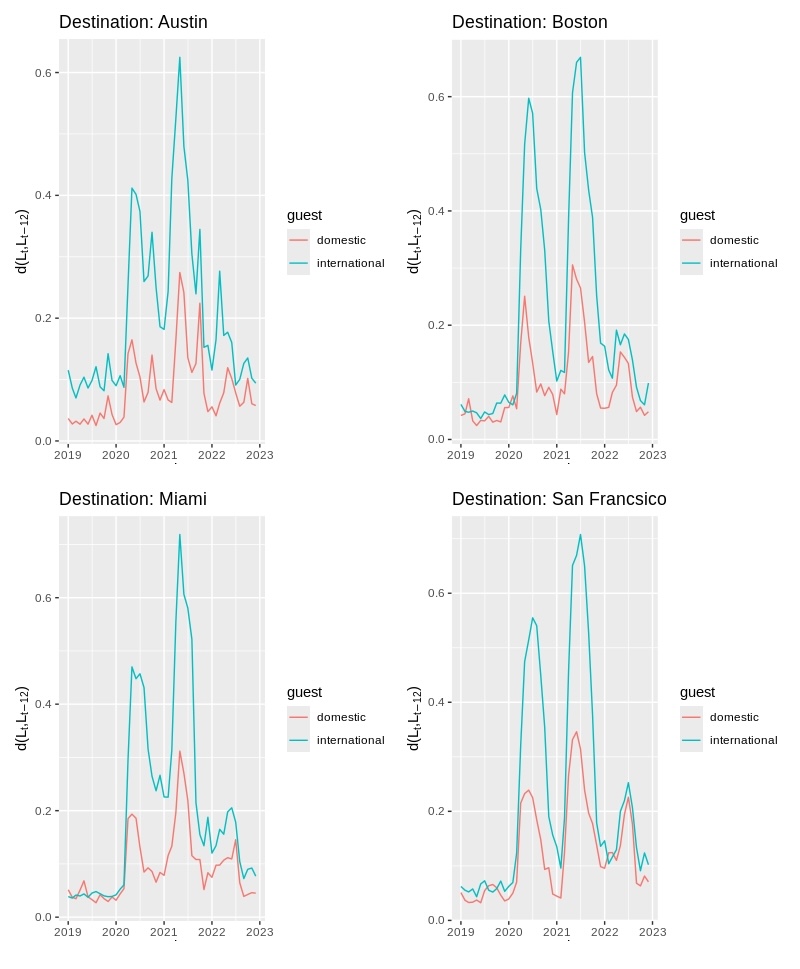}
    \caption{Normalized L1 divergence in \textbf{destination} lead times for Airbnb bookings, taking the previous year as the baseline year.}
        \label{l1_yoy_destination}
\end{figure}

\begin{figure}
    \includegraphics[scale=.85]{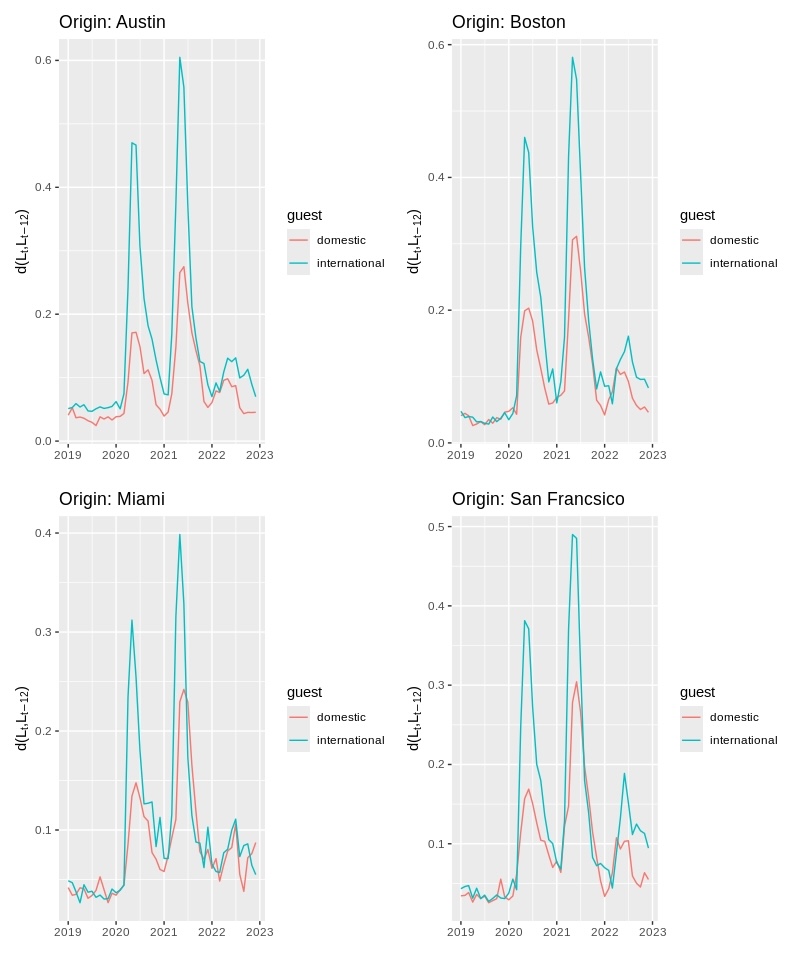}
    \caption{Normalized L1 divergence in \textbf{origin} lead times for Airbnb bookings, taking the previous year as the baseline year.}
        \label{l1_yoy_origin}
\end{figure}

Year-over-year divergences (Figures~\ref{l1_yoy_destination}--\ref{l1_yoy_origin}) highlight temporal shifts \emph{within} the pandemic period. For domestic travel (red lines), $l^1_{\YoY,t}$ often surged in 2021 as the industry began recovering, then tapered in 2022. In San Francisco, the median $l^1_{\YoY,t}$ for domestic destination bookings jumped from 0.04 in 2019 to 0.12 in 2020 and 0.19 in 2021, mirroring the transitions seen in the simpler metrics but adding clarity on \emph{how much} of the distribution had changed. Meanwhile, in Miami, the domestic origin year-over-year divergence rose from 0.04 in 2019 to 0.09 in 2020 and 0.13 in 2021, underscoring ongoing flux.

For international bookings (blue lines), peaks occurred in either 2020 or 2021. The median $l^1_{\YoY,t}$ for bookings from Miami, for instance, reached 0.13 in 2020 before easing slightly; in Boston, the international destination divergence swelled from 0.05 in 2019 to 0.33 in 2020 and 0.39 in 2021. Although these values generally drifted downward in 2022, they remained well above pre-pandemic norms, reflecting prolonged turbulence in international travel. Indeed, the “two-phase” anomaly—where 2020 itself becomes a new, unconventional baseline—left many city pairs experiencing persistently high year-over-year divergences in 2021.

\clearpage

\subsubsection{Correlation Analysis of Lead Time Divergences}

The correlations between domestic and international $l^1$ divergences across cities and over time  are shown in Supplementary Figures \ref{yo18_heatmap} and \ref{yoy_heatmap}. High correlations within the same travel type and direction suggest consistent behavioral shifts, while lower correlations between different travel types indicate distinct factors influencing domestic and international booking patterns.

In Miami, there was a strong correlation between domestic and international divergences, suggesting that local conditions similarly affected both travel segments. For instance, Miami's domestic origin $l^1_{\YoYeightteen,t}$ (bottom-left panel of Supplementary figure \ref{yo18_heatmap}) had a correlation of 0.83 with its international destination divergence. In contrast, San Francisco (bottom-right panel of Supplementary figure \ref{yo18_heatmap}) exhibited lower correlations between domestic and international travel, implying that different factors influenced booking behaviors in these segments.

Lag effects were also observed, particularly in Austin, where domestic trends influenced international trends with a delay (correlation of 0.76 in $l^1_{\YoYeightteen,t}$). Consistently high correlations within categories across cities indicate a degree of interconnectedness in travel patterns, with domestic travel patterns significantly influencing international travel in some cases.

\subsubsection{Seasonal-Trend Decomposition (STL) Analysis}

The Seasonal-Trend decomposition using LOESS (STL) on the normalized L1 distance time series for both the baseline year 2018 ($l^1_{\YoYeightteen,t}$) and the year-over-year ($l^1_{\YoY,t}$) divergences are shown in Supplementary Figures \ref{stl_yo2018_destination}--\ref{stl_yoy_origin}. The trend components from the STL give insights into the persistent changes in booking lead time distributions over time.

For the $l^1_{\YoYeightteen,t}$ divergences (Supplementary Figures \ref{stl_yo2018_destination}-\ref{stl_yo2018_origin}), the trend components for both domestic and international travel peaked around mid-2020 to mid-2021, corresponding with the peak impact of the COVID-19 pandemic on global travel. These trend components do not return to zero by 2022, indicating that booking lead time distributions have not reverted to their pre-pandemic (2018) patterns. For example, the trend components for domestic origin divergences in Boston and San Francisco remain elevated going into 2023, suggesting sustained shifts in booking behaviors in these markets. This persistent divergence implies that travelers continue to book accommodations with lead times significantly different from those in 2018, reflecting lasting changes in planning behavior possibly due to altered risk perceptions, remote work arrangements, or new travel preferences.

In contrast, the $l^1_{\YoY,t}$ divergences (Supplementary Figures \ref{stl_yoy_destination}-\ref{stl_yoy_origin}) show a different pattern. For domestic travel, the trend components generally decrease towards pre-pandemic levels by 2022, indicating that year-over-year changes in booking behaviors are stabilizing. However, the divergences remain higher than zero, suggesting that while the rate of change is normalizing, booking patterns have not fully returned to previous norms. For international origin travel, the trend components show an acceleration of divergence going into 2023. This indicates that international booking behaviors continue to change significantly compared to the immediate past year.

\section{Discussion}
\label{sec:discussion}

\subsection{Key Insights on Lead-Time Distributions}
Our analysis shows that the COVID-19 pandemic triggered marked changes in Airbnb booking lead times, often in ways that simple mean or median metrics would fail to capture. By studying the \textit{entire} distribution of lead times through the normalized L1 distance, we revealed substantial shifts, including cases where domestic travel in certain cities (e.g., Austin, Miami) rebounded more quickly, whereas others (e.g., Boston, San Francisco) continued to diverge from 2018 baselines. While these differences could reflect variations in local economic composition, traveler preferences, or policy approaches \citep{sigala2020tourism}, our data do not directly measure the causal impact of local regulations. Rather, the persistent L1 divergences remind us that forecasting models relying heavily on historical norms may underperform when external shocks alter traveler behavior in fundamental ways.

\subsubsection{Distribution-Wide Reallocation of Lead Times}
Several city-year examples illustrate how L1 can capture internal shifts that remain obscured by stable or modest changes in averages. For instance, in San Francisco’s domestic travel during 2020, the mean lead time rose by only about 3\% (see Table~\ref{yoyeighteendomestic}), yet the mean $l^1_{\YoYeightteen,t}$ exceeded 0.16 (Table~\ref{l1_domestic_2018}), implying that a notable fraction of bookings migrated to the extremes (very short or very long lead times). Likewise, destination Miami’s domestic bookings in 2020 showed a mean that was nearly unchanged from 2018, yet the L1 distance spiked to around 0.11, indicating tail-driven reallocation. And in origin Boston’s domestic market in 2020, the median increased 3\% from the previous year while the median $l^1_{\YoY,t}$ was .11 suggesting that while many travelers booked further out, a subset planned well in advance or much earlier than they did in the previous year, preserving considerable spread overall. In each case, L1 reveals the depth of these distributional shifts, underscoring how stable or even partially “recovered” means and medians can mask significant reconfigurations of booking horizons.

\subsection{Implications for Forecasting and Revenue Management}
In practical terms, elevated L1 distances highlight situations where the historical “booking curve” may be a poor guide to current or future patterns. Revenue managers adopting naive pickup methods could systematically under- or over-predict demand if more guests begin booking at very short notice, or conversely, further in advance. Mitigation strategies include:
\begin{itemize}
    \item \textbf{Recalibrated Forecasts:} Incorporate distribution-aware metrics into forecast models, periodically re-estimating how booking timelines have shifted.
    \item \textbf{Scenario Analyses:} Use the L1 distance as a red flag to explore best- and worst-case demand scenarios, particularly when pivoting to new pricing or marketing campaigns.
    \item \textbf{Monitoring International vs.\ Domestic Patterns:} As our results illustrate, cross-border travel can diverge more drastically from historical patterns, warranting special attention to the L1 divergences in that segment.
\end{itemize}

\subsection{Forward-Looking Extensions of the L1 Distance}
Although the present study focuses on \emph{final} lead-time distributions, a natural extension is to track how in-progress (partial) bookings compare to the \emph{historical} shape of bookings at the same horizon. Concretely:
\begin{enumerate}
  \item \textbf{Partial Distribution Construction:} For each arrival date, identify what fraction of total bookings (so far) accrued at lead times from $0$ to $H$ days out, where $H$ is a “checkpoint” (e.g., 30 days before arrival).
  \item \textbf{Comparative Baseline:} Compute an average partial distribution at $H$ days out using data from prior years or seasons. This yields a historical reference $L^{(H)}_{\text{hist}}$ to which the current partial distribution $L^{(H)}_t$ can be compared.
  \item \textbf{Partial L1 Metric:} 
    \[
      D_{H}(t) \;=\; \frac{1}{2} \sum_{\delta=0}^{H} |L^{(H)}_t(\delta) - L^{(H)}_{\text{hist}}(\delta)|.
    \]
    A higher $D_{H}(t)$ signals that current booking behavior is unfolding quite differently from the norm.
  \item \textbf{Early Warning:} When $D_{H}(t)$ exceeds a chosen threshold, managers can anticipate a final distribution that may deviate from historical patterns. They might respond by adjusting prices, experimenting with targeted promotions, or revising inventory allocations.
\end{enumerate}
Because these “forward-looking” metrics capture divergences \emph{before} the final booking curve is observed, they could serve as an \emph{early-warning tool} in times of heightened volatility. While we have not implemented this approach in the current analysis, it represents a natural follow-up for practitioners seeking near-real-time anomaly detection.

\subsection{Conclusion and Future Directions}
\label{sec:conclusion_future}

By emphasizing the \emph{full} distribution of lead times, this study demonstrates how the COVID-19 pandemic produced lasting changes that are not fully captured by traditional statistics such as the mean or median. In revealing reallocation among short-, medium-, and long-lead segments, the normalized L1 distance emerges as a valuable diagnostic tool for detecting and quantifying booking-behavior shifts. These insights are particularly relevant to industry stakeholders looking to adapt forecasting and operational strategies to volatile travel markets.

Despite this contribution, our analysis is subject to several limitations. First, we focused on four major U.S.\ cities using data from a single accommodation platform. Results may not generalize to all markets or lodging segments, particularly those with smaller or more heterogeneous demand bases. Second, we did not isolate other potential drivers of lead-time divergence, such as flight capacity, local pandemic severity, or remote-work policies. Incorporating these factors might clarify whether certain market segments or traveler demographics are more prone to last-minute shifts.

Building on these insights, future research could broaden the dataset to include additional destinations or compare multiple lodging types. Methodologically, integrating the L1 distance into machine learning or advanced time-series models could improve forecast robustness, especially under fluctuating conditions. Another promising direction involves piloting a \emph{forward-looking} or partial-L1 approach in real-world revenue management systems, where even small early warnings about distributional shifts can yield significant benefits in pricing, marketing, and inventory controls.


\section*{Data Availability}
The primary dataset used in our study, Lead Times in Flux: Analyzing Airbnb Dynamics During Global Upheavals (2018-2022), is not publicly available due to confidentiality constraints.

\bibliographystyle{chicago}
\bibliography{references}

\newpage
\newpage
\section*{Supplementary Materials}
\setcounter{figure}{0}
\renewcommand{\thefigure}{S\arabic{figure}}
\setcounter{table}{0}
\renewcommand{\thetable}{S\arabic{table}}

\subsection{Derivation of the Forecast Error Bound}
\label{appendix:derivation}

We aim to derive the bound:

\[
|\epsilon| \leq \frac{2D \left( 1 - \frac{\Delta}{\Delta_{\text{max}}} \right)}{C_{\text{hist}}(\Delta)}
\]

\subsection*{Definitions}

Let:

\begin{itemize}
    \item \( L_{\text{hist}}(k) \) and \( L_{\text{actual}}(k) \) be the historical and actual lead time distributions.
    \item \( C_{\text{hist}}(\Delta) = \sum_{k=1}^{\Delta} L_{\text{hist}}(k) \) and \( C_{\text{actual}}(\Delta) = \sum_{k=1}^{\Delta} L_{\text{actual}}(k) \) be their cumulative distributions.
    \item \( D = \frac{1}{2} \sum_{k=1}^{\Delta_{\text{max}}} |L_{\text{hist}}(k) - L_{\text{actual}}(k)| \) be the normalized L1 distance.
    \item \( \Delta_\text{max} \) be the maximum lead time.
\end{itemize}

The \textbf{relative forecast error} \( \epsilon \) is defined as:

\[
\epsilon = \frac{\hat{B}_{\text{total}} - B_{\text{total}}}{B_{\text{total}}}
\]

where:

\begin{itemize}
    \item \( \hat{B}_{\text{total}} = \frac{B_{\text{obs}}}{C_{\text{hist}}(\Delta)} \) is the forecasted total bookings using the pickup method.
    \item \( B_{\text{obs}} = C_{\text{actual}}(\Delta) B_{\text{total}} \) is the actual number of bookings observed up to lead time \( \Delta \).
\end{itemize}

\subsection*{Derivation}

Substitute \( \hat{B}_{\text{total}} \) and \( B_{\text{obs}} \) into \( \epsilon \):

\[
\epsilon = \frac{\frac{B_{\text{obs}}}{C_{\text{hist}}(\Delta)} - B_{\text{total}}}{B_{\text{total}}} = \frac{\frac{C_{\text{actual}}(\Delta) B_{\text{total}}}{C_{\text{hist}}(\Delta)} - B_{\text{total}}}{B_{\text{total}}}
\]

Simplify:

\[
\epsilon = \frac{C_{\text{actual}}(\Delta)}{C_{\text{hist}}(\Delta)} - 1 = \frac{C_{\text{actual}}(\Delta) - C_{\text{hist}}(\Delta)}{C_{\text{hist}}(\Delta)}
\]

Take the absolute value:

\[
|\epsilon| = \left| \frac{C_{\text{actual}}(\Delta) - C_{\text{hist}}(\Delta)}{C_{\text{hist}}(\Delta)} \right|
\]

We need to bound \( |C_{\text{actual}}(\Delta) - C_{\text{hist}}(\Delta)| \). Consider:

\[
C_{\text{actual}}(\Delta) - C_{\text{hist}}(\Delta) = \sum_{k=1}^{\Delta}( L_{\text{actual}}(k) - L_{\text{hist}}(k))
\]

Since the total difference between the distributions is \( 2D \):

\[
\sum_{k=1}^{\Delta_\text{max}} |L_{\text{actual}}(k) - L_{\text{hist}}(k)| = 2D
\]

The sum from \( k=1 \) to \( \Delta \) accounts for a fraction \( \frac{\Delta}{\Delta_\text{max}} \) of the total possible divergence (assuming the divergence is uniformly distributed over \( \Delta_\text{max} \)).

Therefore, the maximum cumulative difference up to \( \Delta \) is:

\[
|C_{\text{actual}}(\Delta) - C_{\text{hist}}(\Delta)| \leq 2D \left( 1 - \frac{\Delta}{\Delta_\text{max}} \right)
\]

Substituting back into \( |\epsilon| \):

\[
|\epsilon| \leq \frac{2D \left( 1 - \frac{\Delta}{\Delta_\text{max}} \right)}{C_{\text{hist}}(\Delta)}
\]

This completes the derivation of the forecast error bound.

\subsection*{Implications}
\begin{itemize}
    \item As \( \Delta \) approaches \( \Delta_\text{max} \), \( \left( 1 - \frac{\Delta}{\Delta_\text{max}} \right) \) approaches zero, causing the bound \( |\epsilon| \) to approach zero.
    \item As $D \rightarrow 1$, the bound becomes less informative. 
\end{itemize}

\subsection*{Supplementary Figures and Tables}

\begin{figure}
    \includegraphics[scale=.85]{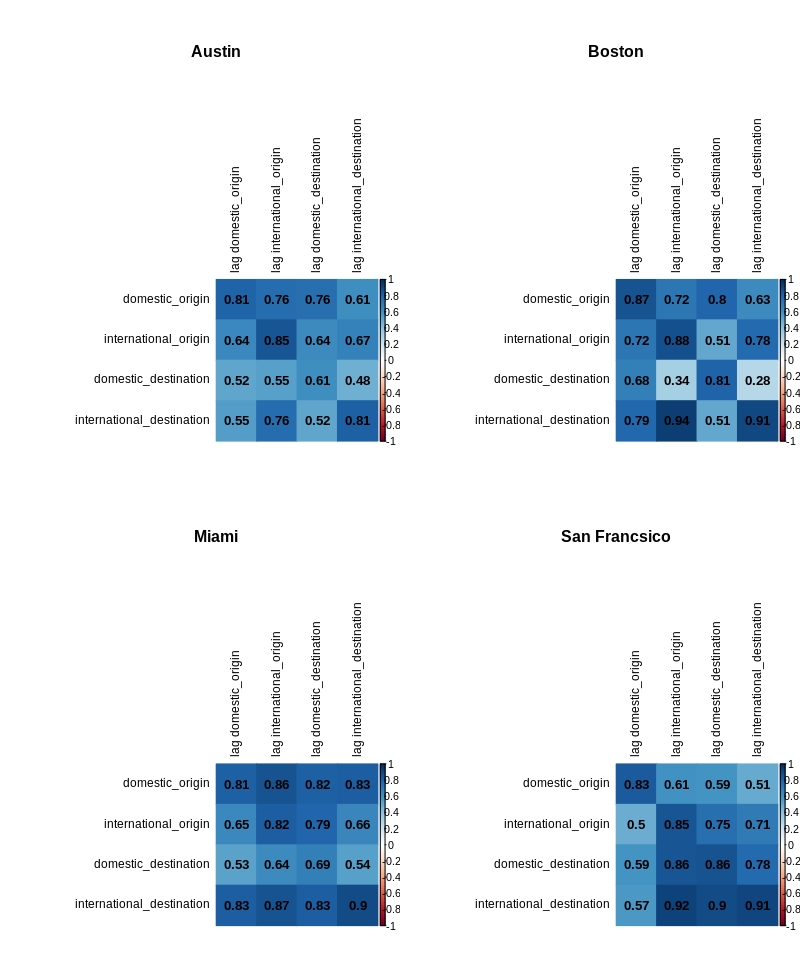}
    \caption{A heatmap displaying the correlations between L1 Yo2018 divergences for Austin, Boston, Miami, and San Francisco and travel types (domestic vs. international), with colder colors indicating stronger positive correlations in divergence behavior.}
        \label{yo18_heatmap}
\end{figure}

\begin{figure}
    \includegraphics[scale=.85]{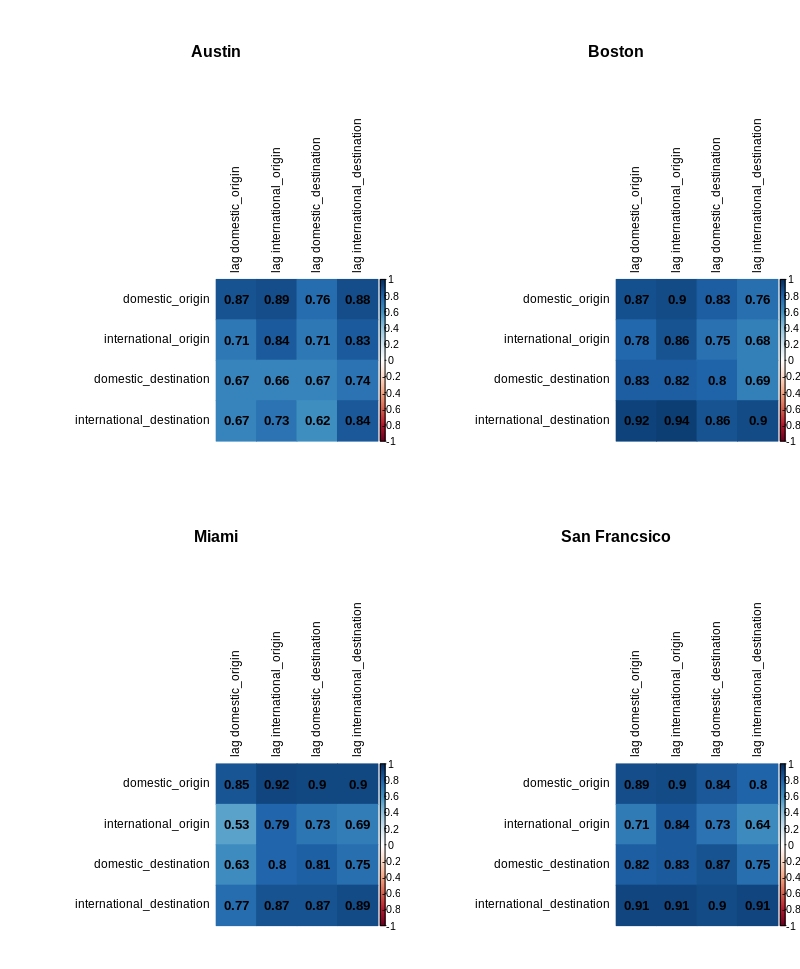}
    \caption{A heatmap displaying the correlations between L1 YoY divergences for Austin, Boston, Miami, and San Francisco and travel types (domestic vs. international), with colder colors indicating stronger positive correlations in divergence behavior.}
        \label{yoy_heatmap}
\end{figure}

\begin{figure}
    \includegraphics[scale=.9]{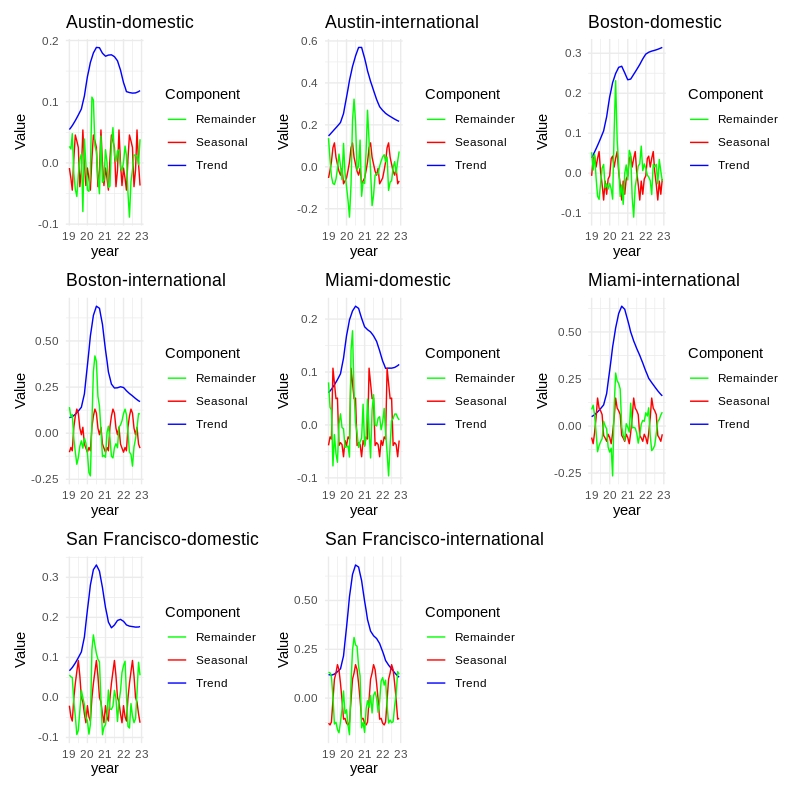}
     \caption{
    Seasonal-Trend decomposition (STL) of the year-over-2018 ($l^1_{\YoYeightteen,t}$) divergence time series for \emph{destination} lead times in four U.S.\ cities (Austin, Boston, Miami, San Francisco). Each panel corresponds to one city-route combination (e.g., Austin-domestic, Boston-international), and the $l^1_{\YoYeightteen,t}$ values are plotted monthly from 2019 to 2023 on the horizontal axis. 
    The \textbf{blue curve} (Trend) shows the longer-term evolution of divergence from the 2018 baseline, the \textbf{red curve} (Seasonal) captures recurring seasonal fluctuations (e.g., holiday or summer travel effects), and the \textbf{green curve} (Remainder) reflects short-term or irregular variations not explained by trend or seasonality. 
    }
        \label{stl_yo2018_destination}
\end{figure}

\begin{figure}
    \includegraphics[scale=.9]{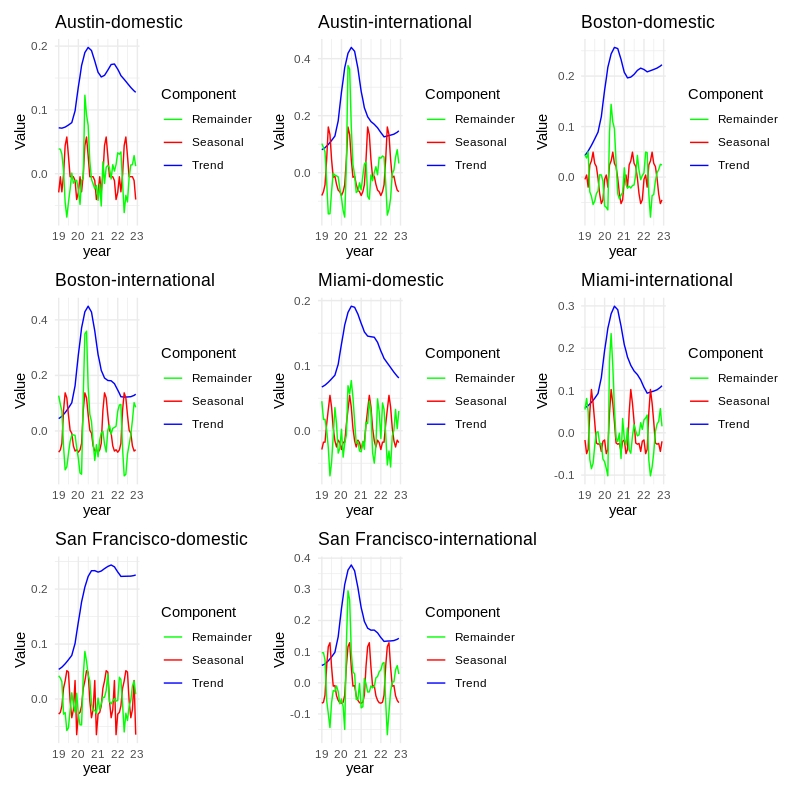}
   \caption{
    Seasonal-Trend decomposition (STL) of the year-over-2018 ($l^1_{\YoYeightteen,t}$) divergence time series for \emph{origin} lead times in four U.S.\ cities (Austin, Boston, Miami, San Francisco). As in Figure~\ref{stl_yo2018_destination}, each panel shows monthly $l^1_{\YoYeightteen,t}$ from 2019 to 2023 for a specific city-route combination (e.g., Miami-international). 
    The decomposition splits the overall divergence (\emph{Value}, on the vertical axis) into: a \textbf{blue Trend line} indicating the long-run trajectory of how far lead-time distributions have shifted from their 2018 baseline; a \textbf{red Seasonal line} capturing repeating patterns driven by annual or event-related cycles; and a \textbf{green Remainder line} revealing short-lived or irregular changes. 
    }
        \label{stl_yo2018_origin}
\end{figure}

\begin{figure}
    \includegraphics[scale=.9]{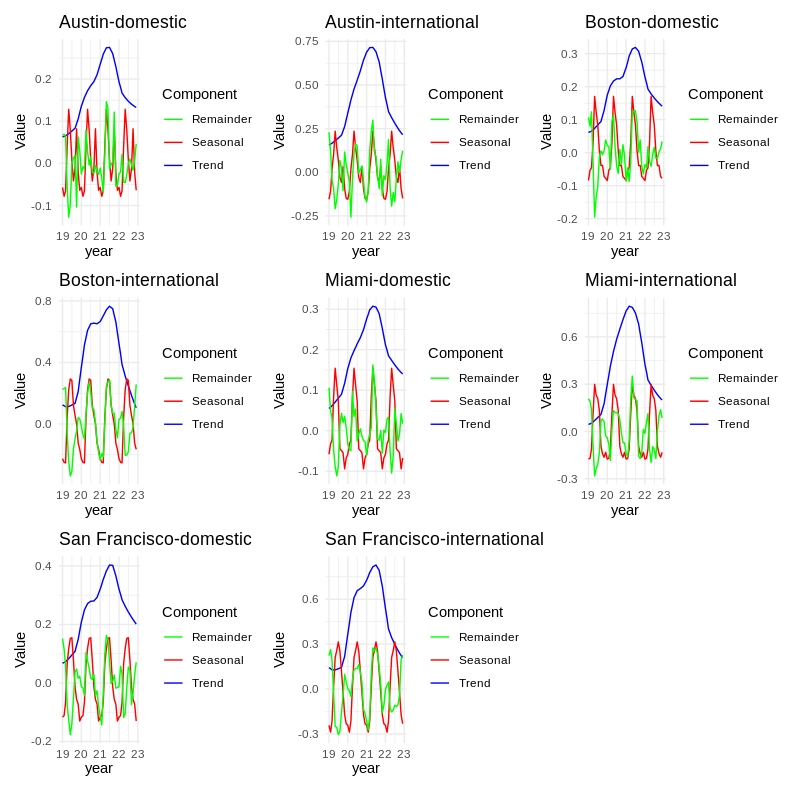}
     \caption{
    Seasonal-Trend decomposition (STL) of the year-over-year ($l^1_{\YoY,t}$) divergence time series for \emph{destination} lead times in four U.S.\ cities (Austin, Boston, Miami, San Francisco). Each panel corresponds to one city-route combination (e.g., Austin-domestic, Boston-international) and spans monthly data from 2019 through 2023. 
    The \textbf{blue Trend curve} tracks the longer-term evolution in how destination lead-time distributions diverge from the previous year, capturing sustained increases or decreases over time. The \textbf{red Seasonal curve} isolates recurring, typically annual patterns (e.g., seasonal peaks tied to holidays or summer travel). The \textbf{green Remainder curve} represents short-lived or irregular fluctuations unaccounted for by the trend and seasonal components. 
    }
        \label{stl_yoy_destination}
\end{figure}

\begin{figure}
    \includegraphics[scale=.9]{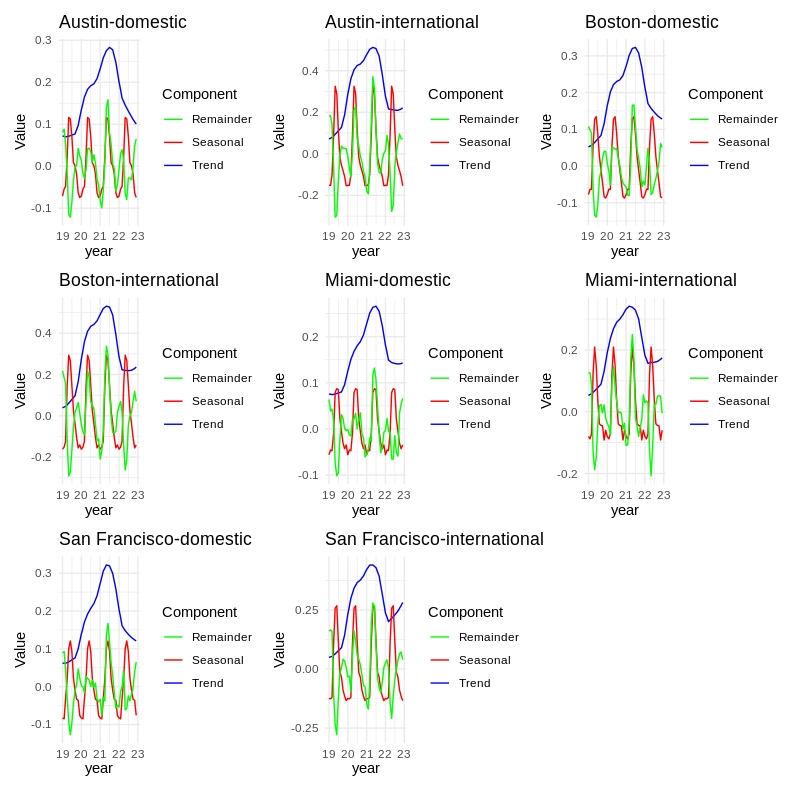}
    \caption{
    Seasonal-Trend decomposition (STL) of the year-over-year ($l^1_{\YoY,t}$) divergence time series for \emph{origin} lead times in four U.S.\ cities (Austin, Boston, Miami, San Francisco). As in Figure~\ref{stl_yoy_destination}, each panel represents a specific city-route pairing over monthly data from 2019 through 2023. 
    The decomposition splits the observed YoY divergence into a \textbf{blue Trend component}, a \textbf{red Seasonal component}, and a \textbf{green Remainder component}. The \textbf{Trend} line reflects longer-term shifts in how far origin lead times deviate from the prior year’s norms, the \textbf{Seasonal} line highlights periodic patterns linked to factors like school breaks or major holidays, and the \textbf{Remainder} captures one-off spikes or dips not explained by either trend or seasonality. 
    }
        \label{stl_yoy_origin}
\end{figure}

\begin{table}[ht]
\centering
\begin{tabular}{rllrlrrr}
  \hline
  city  & year & corridor & mean & median & sd \\ 
  \hline
 Austin   & 2018 & destination & 44 & 28 & 49 \\ 
 Austin   & 2019 & destination & 46 & 29 & 50 \\ 
 Austin   & 2020 & destination & 46 & 28 & 52 \\ 
 Austin   & 2021 & destination & 49 & 34 & 49 \\ 
 Austin   & 2022 & destination & 49 & 30 & 56 \\ 
 Austin   & 2018 & origin & 45 & 28 & 50 \\ 
 Austin   & 2019 & origin & 46 & 29 & 51 \\ 
  Austin   & 2020 & origin & 44 & 27 & 53 \\ 
  Austin   & 2021 & origin & 49 & 33 & 50 \\ 
 Austin   & 2022 & origin & 53 & 34 & 57 \\ 
 Boston   & 2018 & destination & 47 & 29 & 53 \\ 
 Boston   & 2019 & destination & 49 & 30 & 55 \\ 
 Boston   & 2020 & destination & 59 & 35 & 65 \\ 
 Boston   & 2021 & destination & 57 & 40 & 55 \\ 
 Boston   & 2022 & destination & 65 & 46 & 63 \\ 
  Boston   & 2018 & origin & 52 & 32 & 57 \\ 
 Boston   & 2019 & origin & 54 & 34 & 59 \\ 
 Boston   & 2020 & origin & 56 & 34 & 62 \\ 
 Boston   & 2021 & origin & 59 & 41 & 57 \\ 
 Boston   & 2022 & origin & 67 & 45 & 68 \\ 
 Miami   & 2018 & destination & 39 & 24 & 46 \\ 
  Miami   & 2019 & destination &  40 & 24 & 46 \\  
 Miami   & 2020 & destination & 39 & 22 & 49 \\ 
 Miami   & 2021 & destination & 39 & 26 & 43 \\ 
 Miami   & 2022 & destination & 43 & 26 & 51 \\ 
 Miami   & 2018 & origin & 35 & 20 & 44 \\ 
 Miami   & 2019 & origin & 36 & 19 & 46 \\ 
Miami   & 2020 & origin & 33 & 17 & 45 \\ 
 Miami   & 2021 & origin & 39 & 23 & 46 \\ 
 Miami   & 2022 & origin & 40 & 21 & 52 \\ 
 San Francisco   & 2018 & destination & 52 & 37 & 50 \\ 
 San Francisco   & 2019 & destination & 52 & 36 & 52 \\ 
 San Francisco   & 2020 & destination & 54 & 32 & 58 \\ 
 San Francisco   & 2021 & destination & 49 & 36 & 47 \\ 
 San Francisco   & 2022 & destination & 61 & 46 & 55 \\ 
 San Francisco   & 2018 & origin & 42 & 27 & 47 \\ 
 San Francisco   & 2019 & origin & 45 & 28 & 50 \\ 
San Francisco  & 2020 & origin & 43 & 28 & 48 \\ 
 San Francisco   & 2021 & origin & 50 & 36 & 47 \\ 
 San Francisco   & 2022 & origin & 55 & 37 & 56 \\ 
   \hline
\end{tabular}
\caption{Airbnb summary lead times \textbf{domestic}: Summary statistics for domestic Airbnb bookings from 2018 to 2022, including mean, median, and standard deviation of lead times across four U.S. cities (Austin, Boston, Miami, and San Francisco). Origin/destination indicates whether the booking was from/to the city. \label{yearly_summary_domestic}}
\end{table}
\begin{table}[ht]
\centering
\begin{tabular}{rllrlrrr}
  \hline
  city & year & corridor & mean & median & sd \\ 
  \hline
 Austin   & 2018 & destination & 60 & 41 & 60 \\ 
 Austin   & 2019 & destination & 61 & 42 & 60 \\ 
 Austin   & 2020 & destination & 77 & 52 & 74 \\ 
 Austin   & 2021 & destination & 48 & 31 & 50 \\ 
 Austin   & 2022 & destination & 50 & 32 & 53 \\ 
 Austin   & 2018 & origin & 56 & 38 & 57 \\ 
 Austin   & 2019 & origin & 58 & 38 & 60 \\ 
 Austin   & 2020 & origin & 71 & 43 & 74 \\ 
 Austin   & 2021 & origin & 50 & 32 & 54 \\ 
 Austin   & 2022 & origin & 60 & 41 & 61 \\ 
 Boston   & 2018 & destination & 61 & 43 & 58 \\ 
 Boston   & 2019 & destination & 64 & 45 & 62 \\ 
 Boston   & 2020 & destination & 95 & 74 & 80 \\ 
 Boston   & 2021 & destination & 64 & 46 & 60 \\ 
  Boston   & 2022 & destination & 70 & 54 & 62 \\ 
 Boston   & 2018 & origin & 57 & 39 & 58 \\ 
 Boston   & 2019 & origin & 59 & 40 & 59 \\ 
 Boston   & 2020 & origin & 72 & 45 & 73 \\ 
 Boston   & 2021 & origin & 50 & 32 & 55 \\ 
 Boston   & 2022 & origin & 61 & 42 & 62 \\ 
Miami   & 2018 & destination & 61 & 43 & 59 \\ 
 Miami   & 2019 & destination & 61 & 42 & 60 \\ 
 Miami   & 2020 & destination & 78 & 54 & 75 \\ 
Miami   & 2021 & destination & 41 & 26 & 50 \\ 
 Miami   & 2022 & destination & 50 & 32 & 55 \\ 
  Miami   & 2018 & origin & 45 & 27 & 51 \\ 
 Miami   & 2019 & origin & 45 & 26 & 52 \\ 
  Miami   & 2020 & origin & 48 & 25 & 60 \\ 
   Miami   & 2021 & origin & 39 & 23 & 47 \\ 
   Miami   & 2022 & origin & 46 & 27 & 53 \\ 
   San Francisco   & 2018 & destination & 67 & 49 & 60 \\ 
   San Francisco   & 2019 & destination & 69 & 50 & 62 \\ 
   San Francisco   & 2020 & destination & 93 & 71 & 79 \\ 
  San Francisco   & 2021 & destination & 55 & 37 & 57 \\ 
  San Francisco   & 2022 & destination & 64 & 48 & 56 \\ 
   San Francisco   & 2018 & origin & 50 & 34 & 53 \\ 
   San Francisco   & 2019 & origin & 53 & 35 & 55 \\ 
   San Francisco   & 2020 & origin & 59 & 35 & 65 \\ 
   San Francisco   & 2021 & origin & 44 & 29 & 49 \\ 
   San Francisco   & 2022 & origin & 56 & 38 & 57 \\ 
   \hline
\end{tabular}
\caption{Airbnb summary lead times \textbf{international}: Summary statistics for international Airbnb bookings from 2018 to 2022, including mean, median, and standard deviation of lead times across four U.S. cities (Austin, Boston, Miami, and San Francisco). Origin/destination indicates whether the booking was from/to the city. \label{yearly_summary_international}}
\end{table}

\begin{table}[ht]
\centering
\begin{tabular}{rllrlrrr}
  \hline
  city  & year & corridor & mean & median & sd \\ 
  \hline
Austin &   2019 & destination & 1.036 & 1.040 & 1.025 \\ 
 Austin &   2020& destination & 0.998 & 0.962 & 1.041 \\ 
 Austin &   2021 & destination & 1.078 & 1.204 & 0.951 \\ 
  Austin &   2022 & destination & 1.007 & 0.897 & 1.125 \\ 
 Austin &   2019 & origin & 1.022 & 1.028 & 1.020 \\ 
 Austin &   2020 & origin & 0.961 & 0.911 & 1.026 \\ 
 Austin &   2021 & origin & 1.096 & 1.251 & 0.958 \\ 
   Austin &   2022 & origin & 1.097 & 1.037 & 1.141 \\ 
Boston &   2019 & destination & 1.043 & 1.038 & 1.038 \\ 
   Boston &   2020 & destination & 1.198 & 1.164 & 1.178 \\ 
   Boston &   2021 & destination & 0.960 & 1.153 & 0.848 \\ 
 Boston &   2022 & destination & 1.145 & 1.131 & 1.142 \\ 
 Boston &   2019 & origin & 1.046 & 1.057 & 1.038 \\ 
 Boston &   2020 & origin & 1.021 & 0.977 & 1.049 \\ 
 Boston &   2021 & origin & 1.064 & 1.228 & 0.927 \\ 
 Boston &   2022& origin & 1.132 & 1.084 & 1.178 \\ 
   Miami &   2019 & destination & 1.039 & 1.043 & 1.022 \\ 
 Miami &   2020 & destination & 0.973 & 0.903 & 1.038 \\ 
  Miami &   2021 & destination & 0.999 & 1.165 & 0.888 \\ 
Miami &   2022 & destination & 1.109 & 0.990 & 1.191 \\ 
 Miami &   2019& origin & 1.006 & 0.972 & 1.049 \\ 
 Miami &   2020 & origin & 0.928 & 0.913 & 0.977 \\ 
  Miami &   2021 & origin & 1.169 & 1.349 & 1.020 \\ 
 Miami &   2022 & origin & 1.034 & 0.910 & 1.120 \\ 
 San Francisco &   2019 & destination & 1.000 & 0.969 & 1.032 \\ 
San Francisco &   2020 & destination & 1.023 & 0.902 & 1.130 \\ 
San Francisco &   2021 & destination & 0.918 & 1.105 & 0.799 \\ 
  San Francisco &   2022 & destination & 1.245 & 1.285 & 1.184 \\ 
  San Francisco &   2019 & origin & 1.052 & 1.048 & 1.053 \\ 
   San Francisco &   2020 & origin & 0.948 & 0.979 & 0.957 \\ 
San Francisco &   2021 & origin & 1.176 & 1.310 & 0.987 \\ 
  San Francisco &   2022 & origin & 1.104 & 1.024 & 1.194 \\ 
   \hline
\end{tabular}
\caption{Year-over-Year (YoY) metrics for \textbf{domestic} travel. The table presents the mean, median, and standard deviation (sd) of lead times for Airbnb bookings across all months for Austin, Boston, Miami, and San Francisco from 2019 to 2022.  \label{yoydomestic}}
\end{table}

\begin{table}[ht]
\centering
\begin{tabular}{rllrlrrr}
  \hline
  city  & year & corridor & mean & median & sd \\ 
  \hline
Austin  & 2019 & destination & 1.008 & 1.013 & 1.004 \\ 
 Austin  & 2020 & destination & 1.271 & 1.245 & 1.225 \\ 
  Austin  & 2021 & destination & 0.617 & 0.593 & 0.681 \\ 
 Austin  & 2022 & destination & 1.048 & 1.037 & 1.059 \\ 
 Austin  & 2019 & origin & 1.029 & 1.009 & 1.052 \\ 
 Austin  & 2020 & origin & 1.227 & 1.115 & 1.239 \\ 
  Austin  & 2021 & origin & 0.697 & 0.757 & 0.723 \\ 
 Austin  & 2022 & origin & 1.205 & 1.254 & 1.129 \\ 
 Boston  & 2019 & destination & 1.057 & 1.049 & 1.065 \\ 
 Boston  & 2020 & destination & 1.482 & 1.659 & 1.297 \\ 
 Boston  & 2021 & destination & 0.676 & 0.626 & 0.748 \\ 
 Boston  & 2022 & destination & 1.094 & 1.154 & 1.038 \\ 
 Boston  & 2019 & origin & 1.019 & 1.014 & 1.022 \\ 
 Boston  & 2020 & origin & 1.233 & 1.137 & 1.242 \\ 
 Boston  & 2021 & origin & 0.697 & 0.717 & 0.757 \\ 
 Boston  & 2022 & origin & 1.214 & 1.287 & 1.113 \\ 
  Miami  & 2019 & destination & 1.006 & 0.991 & 1.025 \\ 
 Miami  & 2020 & destination & 1.281 & 1.281 & 1.238 \\ 
 Miami  & 2021 & destination & 0.526 & 0.477 & 0.666 \\ 
 Miami  & 2022 & destination & 1.220 & 1.230 & 1.113 \\ 
 Miami  & 2019 & origin & 0.999 & 0.985 & 1.010 \\ 
  Miami  & 2020 & origin & 1.082 & 0.958 & 1.159 \\ 
  Miami  & 2021 & origin & 0.801 & 0.916 & 0.787 \\ 
 Miami  & 2022 & origin & 1.179 & 1.179 & 1.118 \\ 
  San Francisco  & 2019 & destination & 1.020 & 1.013 & 1.040 \\ 
 San Francisco  & 2020 & destination & 1.361 & 1.423 & 1.262 \\ 
 San Francisco  & 2021 & destination & 0.584 & 0.521 & 0.720 \\ 
 San Francisco  & 2022 & destination & 1.165 & 1.298 & 0.995 \\ 
 San Francisco  & 2019 & origin & 1.041 & 1.029 & 1.050 \\ 
 San Francisco  & 2020 & origin & 1.123 & 1.006 & 1.173 \\ 
  San Francisco  & 2021 & origin & 0.751 & 0.836 & 0.757 \\ 
San Francisco  & 2022 & origin & 1.255 & 1.306 & 1.158 \\ 
   \hline
\end{tabular}
\caption{Year-over-Year (YoY) metrics for \textbf{international} travel. The table presents the mean, median, and standard deviation (sd) of lead times for Airbnb bookings across all months for Austin, Boston, Miami, and San Francisco from 2019 to 2022.  \label{yoyinternational}}
\end{table}

\begin{table}[ht]
\centering
\begin{tabular}{rllrlrrr}
  \hline
  city &  year & corridor & mean & median & sd \\ 
  \hline
 Austin   & 2019 & destination & 1.036 & 1.040 & 1.025 \\ 
   Austin   & 2020 & destination & 1.034 & 1.000 & 1.067 \\ 
   Austin   & 2021 & destination & 1.114 & 1.204 & 1.015 \\ 
   Austin   & 2022 & destination & 1.122 & 1.080 & 1.141 \\ 
   Austin   & 2019 & origin & 1.022 & 1.028 & 1.020 \\ 
   Austin   & 2020 & origin & 0.982 & 0.937 & 1.046 \\ 
   Austin   & 2021 & origin & 1.076 & 1.172 & 1.002 \\ 
   Austin   & 2022 & origin & 1.180 & 1.215 & 1.143 \\ 
 Boston   & 2019 & destination & 1.043 & 1.038 & 1.038 \\ 
   Boston   & 2020 & destination & 1.250 & 1.208 & 1.223 \\ 
   Boston   & 2021 & destination & 1.200 & 1.393 & 1.036 \\ 
 Boston   & 2022 & destination & 1.374 & 1.576 & 1.184 \\ 
 Boston   & 2019 & origin & 1.046 & 1.057 & 1.038 \\ 
 Boston   & 2020 & origin & 1.068 & 1.033 & 1.088 \\ 
  Boston   & 2021 & origin & 1.137 & 1.268 & 1.009 \\ 
Boston   & 2022 & origin & 1.287 & 1.374 & 1.188 \\ 
 Miami   & 2019 & destination & 1.039 & 1.043 & 1.022 \\ 
Miami   & 2020 & destination & 1.011 & 0.942 & 1.061 \\ 
 Miami   & 2021 & destination & 1.010 & 1.097 & 0.941 \\ 
 Miami   & 2022 & destination & 1.120 & 1.085 & 1.121 \\ 
 Miami   & 2019 & origin & 1.006 & 0.972 & 1.049 \\ 
 Miami   & 2020 & origin & 0.934 & 0.888 & 1.025 \\ 
  Miami   & 2021 & origin & 1.092 & 1.197 & 1.045 \\ 
Miami   & 2022 & origin & 1.128 & 1.089 & 1.170 \\ 
 San Francisco   & 2019 & destination & 1.000 & 0.969 & 1.032 \\ 
 San Francisco   & 2020 & destination & 1.023 & 0.875 & 1.167 \\ 
San Francisco   & 2021 & destination & 0.939 & 0.966 & 0.932 \\ 
San Francisco   & 2022 & destination & 1.169 & 1.242 & 1.104 \\ 
  San Francisco   & 2019 & origin & 1.052 & 1.048 & 1.053 \\ 
 San Francisco   & 2020 & origin & 0.998 & 1.027 & 1.007 \\ 
  San Francisco   & 2021 & origin & 1.173 & 1.345 & 0.995 \\ 
 San Francisco   & 2022 & origin & 1.295 & 1.377 & 1.188 \\ 
   \hline
\end{tabular}
\caption{Year-over-Year (YoY) metrics for \textbf{domestic} travel using 2018 as the base year. The table presents the mean, median, and standard deviation (sd) of lead times for Airbnb bookings across all months for Austin, Boston, Miami, and San Francisco from 2019 to 2022.  \label{yoyeighteendomestic}}
\end{table}

\begin{table}[ht]
\centering
\begin{tabular}{rllrlrrr}
  \hline
  city  &year & corridor & mean & median & sd \\ 
  \hline
Austin  & 2019 & destination & 1.008 & 1.013 & 1.004 \\ 
 Austin  & 2020 & destination & 1.281 & 1.261 & 1.230 \\ 
 Austin  & 2021 & destination & 0.791 & 0.748 & 0.838 \\ 
 Austin  & 2022 & destination & 0.829 & 0.776 & 0.887 \\ 
 Austin  & 2019 & origin & 1.029 & 1.009 & 1.052 \\ 
 Austin  & 2020 & origin & 1.263 & 1.125 & 1.304 \\ 
   Austin  & 2021 & origin & 0.880 & 0.851 & 0.942 \\ 
   Austin  & 2022 & origin & 1.060 & 1.068 & 1.064 \\ 
 Boston  & 2019 & destination & 1.057 & 1.049 & 1.065 \\ 
 Boston  & 2020 & destination & 1.567 & 1.741 & 1.381 \\ 
 Boston  & 2021 & destination & 1.060 & 1.090 & 1.033 \\ 
  Boston  & 2022 & destination & 1.159 & 1.257 & 1.072 \\ 
 Boston  & 2019 & origin & 1.019 & 1.014 & 1.022 \\ 
 Boston  & 2020 & origin & 1.256 & 1.153 & 1.270 \\ 
 Boston  & 2021 & origin & 0.876 & 0.827 & 0.962 \\ 
   Boston  & 2022 & origin & 1.063 & 1.064 & 1.070 \\ 
   Miami  & 2019 & destination & 1.006 & 0.991 & 1.025 \\ 
 Miami  & 2020 & destination & 1.288 & 1.270 & 1.269 \\ 
   Miami  & 2021 & destination & 0.678 & 0.605 & 0.845 \\ 
   Miami  & 2022 & destination & 0.827 & 0.744 & 0.940 \\ 
 Miami  & 2019 & origin & 0.999 & 0.985 & 1.010 \\ 
 Miami  & 2020 & origin & 1.081 & 0.944 & 1.170 \\ 
 Miami  & 2021 & origin & 0.865 & 0.864 & 0.921 \\ 
 Miami  & 2022 & origin & 1.020 & 1.019 & 1.029 \\ 
 San Francisco  & 2019 & destination & 1.020 & 1.013 & 1.040 \\ 
 San Francisco  & 2020 & destination & 1.389 & 1.441 & 1.313 \\ 
 San Francisco  & 2021 & destination & 0.811 & 0.751 & 0.946 \\ 
   San Francisco  & 2022 & destination & 0.945 & 0.975 & 0.941 \\ 
   San Francisco  & 2019 & origin & 1.041 & 1.029 & 1.050 \\ 
 San Francisco  & 2020 & origin & 1.169 & 1.034 & 1.232 \\ 
  San Francisco  & 2021 & origin & 0.877 & 0.865 & 0.933 \\ 
 San Francisco  & 2022 & origin & 1.101 & 1.130 & 1.081 \\ 
   \hline
\end{tabular}
\caption{Year-over-Year (YoY) metrics for \textbf{international} travel using 2018 as the base year. The table presents the mean, median, and standard deviation (sd) of lead times for Airbnb bookings across all months for Austin, Boston, Miami, and San Francisco from 2019 to 2022.\label{yoyeighteeninternational}}
\end{table}

\begin{table}[ht]
\centering
\begin{tabular}{rlrlllrrrr}
  \hline
 city & year & corridor & mean & median & q025 & q0975 \\ 
  \hline
Austin & 2019 & destination   &   0.04 & 0.04 & 0.03 & 0.07 \\ 
 Austin & 2020 & destination   &   0.09 & 0.08 & 0.03 & 0.16 \\ 
Austin & 2021 & destination   &   0.13 & 0.12 & 0.05 & 0.27 \\ 
 Austin & 2022 & destination   &   0.07 & 0.06 & 0.04 & 0.11 \\ 
 Austin & 2019 & origin   &   0.04 & 0.04 & 0.03 & 0.05 \\ 
Austin & 2020 & origin   &   0.09 & 0.09 & 0.04 & 0.17 \\ 
 Austin & 2021 & origin   &   0.13 & 0.13 & 0.04 & 0.27 \\ 
 Austin & 2022 & origin   &   0.07 & 0.07 & 0.04 & 0.10 \\ 
 Boston & 2019 & destination   &   0.04 & 0.03 & 0.03 & 0.07 \\ 
 Boston & 2020 & destination   &   0.11 & 0.09 & 0.05 & 0.23 \\ 
 Boston & 2021 & destination   &   0.15 & 0.14 & 0.05 & 0.30 \\ 
 Boston & 2022 & destination   &   0.08 & 0.07 & 0.04 & 0.15 \\ 
 Boston & 2019 & origin   &   0.04 & 0.04 & 0.03 & 0.05 \\ 
 Boston & 2020 & origin   &   0.11 & 0.10 & 0.04 & 0.20 \\ 
 Boston & 2021 & origin   &   0.16 & 0.14 & 0.06 & 0.31 \\ 
 Boston & 2022 & origin   &   0.07 & 0.07 & 0.04 & 0.11 \\ 
 Miami & 2019 & destination   &   0.04 & 0.04 & 0.03 & 0.06 \\ 
 Miami & 2020 & destination   &   0.10 & 0.09 & 0.04 & 0.19 \\ 
 Miami & 2021 & destination   &   0.15 & 0.12 & 0.06 & 0.30 \\ 
   Miami & 2022 & destination   &   0.08 & 0.09 & 0.04 & 0.14 \\ 
Miami & 2019 & origin   &   0.04 & 0.04 & 0.03 & 0.05 \\ 
  Miami & 2020 & origin   &   0.09 & 0.08 & 0.04 & 0.14 \\ 
Miami & 2021 & origin   &   0.13 & 0.10 & 0.06 & 0.24 \\ 
 Miami & 2022 & origin   &   0.07 & 0.07 & 0.04 & 0.10 \\ 
San Francisco & 2019 & destination   &   0.05 & 0.04 & 0.03 & 0.07 \\ 
 San Francisco & 2020 & destination   &   0.14 & 0.12 & 0.04 & 0.24 \\ 
 San Francisco & 2021 & destination   &   0.19 & 0.19 & 0.04 & 0.34 \\ 
 San Francisco & 2022 & destination   &   0.12 & 0.12 & 0.06 & 0.22 \\ 
 San Francisco & 2019 & origin   &   0.03 & 0.03 & 0.03 & 0.05 \\ 
  San Francisco & 2020 & origin   &   0.10 & 0.10 & 0.03 & 0.17 \\ 
   San Francisco & 2021 & origin   &   0.16 & 0.13 & 0.06 & 0.30 \\ 
   San Francisco & 2022 & origin   &   0.07 & 0.06 & 0.04 & 0.11 \\ 
   \hline
\end{tabular}

\caption{$l^1_{\YoY,t}$: Year-over-Year normalized L1 divergence for \textbf{domestic} travel lead times, comparing the distributions of bookings from 2019 to 2022 for Austin, Boston, Miami, and San Francisco.  \label{l1_domestic_yoy}}
\end{table}

\begin{table}[ht]
\centering
\begin{tabular}{rlrlllrrrr}
  \hline
  city & year & corridor  & mean & median & q025 & q0975 \\ 
  \hline
Austin & 2019 & destination   & 0.10 & 0.09 & 0.07 & 0.14 \\ 
 Austin & 2020 & destination   & 0.25 & 0.26 & 0.09 & 0.41 \\ 
 Austin & 2021 & destination   & 0.34 & 0.32 & 0.15 & 0.60 \\ 
  Austin & 2022 & destination   & 0.14 & 0.13 & 0.09 & 0.25 \\ 
 Austin & 2019 & origin   & 0.05 & 0.05 & 0.05 & 0.06 \\ 
 Austin & 2020 & origin   & 0.21 & 0.17 & 0.05 & 0.47 \\ 
   Austin & 2021 & origin   & 0.24 & 0.17 & 0.07 & 0.59 \\ 
 Austin & 2022 & origin   & 0.10 & 0.10 & 0.07 & 0.13 \\ 
 Boston & 2019 & destination   & 0.05 & 0.05 & 0.04 & 0.07 \\ 
  Boston & 2020 & destination   & 0.31 & 0.33 & 0.06 & 0.59 \\ 
  Boston & 2021 & destination   & 0.37 & 0.39 & 0.11 & 0.67 \\ 
  Boston & 2022 & destination   & 0.13 & 0.13 & 0.06 & 0.19 \\ 
   Boston & 2019 & origin   & 0.04 & 0.04 & 0.03 & 0.05 \\ 
 Boston & 2020 & origin   & 0.21 & 0.19 & 0.04 & 0.45 \\ 
 Boston & 2021 & origin   & 0.25 & 0.17 & 0.07 & 0.57 \\ 
 Boston & 2022 & origin   & 0.11 & 0.10 & 0.07 & 0.15 \\ 
 Miami & 2019 & destination   & 0.04 & 0.04 & 0.04 & 0.05 \\ 
 Miami & 2020 & destination   & 0.28 & 0.28 & 0.04 & 0.47 \\ 
  Miami & 2021 & destination   & 0.37 & 0.27 & 0.14 & 0.69 \\ 
  Miami & 2022 & destination   & 0.13 & 0.13 & 0.07 & 0.20 \\ 
  Miami & 2019 & origin   & 0.04 & 0.04 & 0.03 & 0.05 \\ 
 Miami & 2020 & origin   & 0.14 & 0.13 & 0.04 & 0.30 \\ 
Miami & 2021 & origin   & 0.16 & 0.11 & 0.06 & 0.38 \\ 
 Miami & 2022 & origin   & 0.08 & 0.08 & 0.06 & 0.11 \\ 
 San Francisco & 2019 & destination   & 0.06 & 0.06 & 0.05 & 0.07 \\ 
 San Francisco & 2020 & destination   & 0.32 & 0.34 & 0.06 & 0.55 \\ 
 San Francisco & 2021 & destination   & 0.40 & 0.42 & 0.11 & 0.70 \\ 
 San Francisco & 2022 & destination   & 0.15 & 0.13 & 0.09 & 0.24 \\ 
   San Francisco & 2019 & origin   & 0.04 & 0.03 & 0.03 & 0.05 \\ 
   San Francisco & 2020 & origin   & 0.18 & 0.16 & 0.04 & 0.38 \\ 
 San Francisco & 2021& origin   & 0.21 & 0.13 & 0.07 & 0.49 \\ 
   San Francisco & 2022 & origin   & 0.11 & 0.11 & 0.05 & 0.18 \\ 
   \hline
\end{tabular}
\caption{$l^1_{\YoY,t}$: Year-over-Year normalized L1 divergence for \textbf{international} travel lead times, comparing the distributions of bookings from 2019 to 2022 for Austin, Boston, Miami, and San Francisco.  \label{l1_international_yoy}}

\end{table}

\begin{table}[ht]
\centering
\begin{tabular}{rlrlllrrrr}
  \hline
  city & year & corridor   & mean & median & q025 & q0975 \\ 
  \hline
 Austin & 2019 & destination   &   0.04 & 0.04 & 0.03 & 0.07 \\ 
  Austin & 2020 & destination   &   0.09 & 0.07 & 0.04 & 0.16 \\ 
 Austin & 2021 & destination   &   0.09 & 0.08 & 0.05 & 0.13 \\ 
  Austin & 2022 & destination   &   0.06 & 0.06 & 0.04 & 0.08 \\ 
 Austin & 2019 & origin   &   0.04 & 0.04 & 0.03 & 0.05 \\ 
 Austin & 2020 & origin   &   0.10 & 0.08 & 0.04 & 0.18 \\ 
 Austin & 2021 & origin   &   0.08 & 0.08 & 0.04 & 0.11 \\ 
 Austin & 2022 & origin   &   0.07 & 0.07 & 0.05 & 0.09 \\ 
Boston & 2019 & destination   &   0.04 & 0.03 & 0.03 & 0.07 \\ 
Boston & 2020 & destination   &   0.13 & 0.10 & 0.08 & 0.24 \\ 
 Boston & 2021 & destination   &   0.13 & 0.13 & 0.09 & 0.16 \\ 
 Boston & 2022 & destination   &   0.15 & 0.15 & 0.11 & 0.18 \\ 
   Boston & 2019 & origin   &   0.04 & 0.04 & 0.03 & 0.05 \\ 
 Boston & 2020 & origin   &   0.12 & 0.10 & 0.06 & 0.21 \\ 
 Boston & 2021 & origin   &   0.10 & 0.10 & 0.08 & 0.12 \\ 
 Boston & 2022 & origin   &   0.11 & 0.11 & 0.09 & 0.13 \\ 
 Miami & 2019 & destination   &   0.04 & 0.04 & 0.03 & 0.06 \\ 
 Miami & 2020 & destination   &   0.11 & 0.08 & 0.05 & 0.23 \\ 
  Miami & 2021 & destination   &   0.08 & 0.07 & 0.05 & 0.14 \\ 
 Miami & 2022 & destination   &   0.05 & 0.05 & 0.03 & 0.08 \\ 
Miami & 2019 & origin   &   0.04 & 0.04 & 0.03 & 0.05 \\ 
Miami & 2020 & origin   &   0.09 & 0.08 & 0.05 & 0.15 \\ 
Miami & 2021 & origin   &   0.07 & 0.07 & 0.04 & 0.12 \\ 
 Miami & 2022 & origin   &   0.05 & 0.05 & 0.04 & 0.07 \\ 
 San Francisco & 2019 & destination   &   0.05 & 0.04 & 0.03 & 0.07 \\ 
 San Francisco & 2020 & destination   &   0.16 & 0.17 & 0.06 & 0.26 \\ 
 San Francisco & 2021 & destination   &   0.09 & 0.09 & 0.06 & 0.14 \\ 
  San Francisco & 2022 & destination   &   0.09 & 0.09 & 0.06 & 0.12 \\ 
 San Francisco & 2019 & origin   &   0.03 & 0.03 & 0.03 & 0.05 \\ 
 San Francisco & 2020 & origin   &   0.11 & 0.11 & 0.04 & 0.17 \\ 
 San Francisco & 2021 & origin   &   0.12 & 0.11 & 0.09 & 0.16 \\ 
 San Francisco & 2022 & origin   &   0.11 & 0.12 & 0.09 & 0.14 \\ 
   \hline
\end{tabular}

\caption{$l^1_{\YoYeightteen,t}$: Normalized L1 divergence for \textbf{domestic} travel lead times using 2018 as the baseline year, comparing the distributions of bookings from 2019 to 2022 for Austin, Boston, Miami, and San Francisco.\label{l1_domestic_2018}}
\end{table}

\begin{table}[ht]
\centering
\begin{tabular}{rlrlllrrrr}
  \hline
  city & year & corridor   & mean & median & q025 & q0975 \\ 
  \hline
Austin & 2019 & destination &     0.10 & 0.09 & 0.07 & 0.14 \\ 
Austin & 2020 & destination &     0.25 & 0.24 & 0.09 & 0.43 \\ 
Austin & 2021 & destination &     0.19 & 0.14 & 0.11 & 0.36 \\ 
 Austin & 2022 & destination &     0.12 & 0.11 & 0.08 & 0.18 \\ 
 Austin & 2019 & origin &     0.05 & 0.05 & 0.05 & 0.06 \\ 
 Austin & 2020 & origin &     0.21 & 0.17 & 0.06 & 0.47 \\ 
 Austin & 2021 & origin &     0.09 & 0.08 & 0.05 & 0.15 \\ 
 Austin & 2022 & origin &     0.06 & 0.07 & 0.05 & 0.08 \\ 
 Boston & 2019 & destination &     0.05 & 0.05 & 0.04 & 0.07 \\ 
 Boston & 2020 & destination &     0.32 & 0.34 & 0.08 & 0.60 \\ 
 Boston & 2021 & destination &     0.13 & 0.12 & 0.09 & 0.16 \\ 
 Boston & 2022 & destination &     0.11 & 0.11 & 0.07 & 0.14 \\ 
Boston & 2019 & origin &     0.04 & 0.04 & 0.03 & 0.05 \\ 
 Boston & 2020 & origin &     0.21 & 0.19 & 0.05 & 0.46 \\ 
 Boston & 2021 & origin &     0.09 & 0.08 & 0.05 & 0.14 \\ 
 Boston & 2022 & origin &     0.06 & 0.06 & 0.04 & 0.09 \\ 
 Miami & 2019 & destination &     0.04 & 0.04 & 0.04 & 0.05 \\ 
 Miami & 2020 & destination &     0.28 & 0.29 & 0.05 & 0.47 \\ 
  Miami & 2021 & destination &     0.21 & 0.22 & 0.14 & 0.30 \\ 
 Miami & 2022 & destination &     0.11 & 0.10 & 0.07 & 0.15 \\ 
  Miami & 2019 & origin &     0.04 & 0.04 & 0.03 & 0.05 \\ 
   Miami & 2020 & origin &     0.14 & 0.12 & 0.05 & 0.30 \\ 
  Miami & 2021 & origin &     0.07 & 0.07 & 0.05 & 0.11 \\ 
   Miami & 2022 & origin &     0.05 & 0.05 & 0.04 & 0.06 \\ 
 San Francisco & 2019 & destination &     0.06 & 0.06 & 0.05 & 0.07 \\ 
 San Francisco & 2020 & destination &     0.32 & 0.35 & 0.08 & 0.56 \\ 
   San Francisco & 2021 & destination &     0.16 & 0.15 & 0.08 & 0.25 \\ 
   San Francisco & 2022 & destination &     0.08 & 0.08 & 0.06 & 0.10 \\ 
   San Francisco & 2019& origin &     0.04 & 0.03 & 0.03 & 0.05 \\ 
   San Francisco & 2020 & origin &     0.18 & 0.15 & 0.06 & 0.39 \\ 
   San Francisco & 2021 & origin &     0.08 & 0.07 & 0.04 & 0.13 \\ 
 San Francisco & 2022 & origin &     0.07 & 0.07 & 0.04 & 0.08 \\ 
   \hline
\end{tabular}

\caption{$l^1_{\YoYeightteen,t}$: Normalized L1 divergence for \textbf{international} travel lead times using 2018 as the baseline year, comparing the distributions of bookings from 2019 to 2022 for Austin, Boston, Miami, and San Francisco.\label{l1_international_2018}}
\end{table}

\end{document}